\documentclass[aip,apl,reprint,a4paper,floatfix,amsmath,amssymb,amsfonts,noshowpacs,citeautoscript]{revtex4-2}
\usepackage{newtxtext,newtxmath}
\usepackage[T1]{fontenc}
\usepackage{graphicx}
\usepackage[dvipsnames]{xcolor}
\usepackage{soul} 
\usepackage{floatrow}
\usepackage{siunitx}
\usepackage[
,textwidth=17.5cm
,textheight=23.5cm
,verbose
,pdftex
]{geometry}
\usepackage{xspace}
\usepackage{svg}
\usepackage[pdftex]{hyperref}
\hypersetup{
  pdftitle={Excitonic and deep-level emission from N- and Al-polar homoepitaxial AlN layers grown by molecular beam epitaxy},
	colorlinks,
	citecolor=blue,
	linkcolor=blue
	}

\makeatletter
\newcommand*{\refig}[2]{\hyperref[#1]{\ref*{#1}(#2)}}
\makeatother

\DeclareMathAlphabet{\mathsf}{OT1}{\sfdefault}{m}{n}
\SetMathAlphabet{\mathsf}{bold}{OT1}{\sfdefault}{b}{n}

\begin{document}

\preprint{AIP/123-QED}

\title[Sample title]{Ultrawide bandgap semiconductor heterojunction p-n diodes with distributed polarization doped p-type AlGaN layers on bulk AlN substrates}
\author{Shivali Agrawal}
\email{sa2368@cornell.edu}
\affiliation{%
Department of Chemical and Biomolecular Engineering, Cornell University, Ithaca, New York 14853, USA
}%
\author{Len van Deurzen}
\affiliation{School of Applied and Engineering Physics, Cornell University, Ithaca, New York 14853, USA}

\author{Jimy Encomendero}
\affiliation{Department of Electrical and Computer Engineering, Cornell University, Ithaca, New York 14853, USA}

\author{Joseph E. Dill}
\affiliation{School of Applied and Engineering Physics, Cornell University, Ithaca, New York 14853, USA}

\author{Hsin Wei (Sheena) Huang}
\affiliation{Department of Electrical and Computer Engineering, Cornell University, Ithaca, New York 14853, USA}

\author{Vladimir Protasenko}
\affiliation{Department of Electrical and Computer Engineering, Cornell University, Ithaca, New York 14853, USA}

\author{Huili (Grace) Xing}
\affiliation{Department of Electrical and Computer Engineering, Cornell University, Ithaca, New York 14853, USA}
\affiliation{Department of Materials Science and Engineering, Cornell University, Ithaca, New York 14853, USA}
\affiliation{Kavli Institute at Cornell for Nanoscale Science, Cornell University, Ithaca, New York 14853, USA}

\author{Debdeep Jena}
\affiliation{Department of Electrical and Computer Engineering, Cornell University, Ithaca, New York 14853, USA}
\affiliation{School of Applied and Engineering Physics, Cornell University, Ithaca, New York 14853, USA}
\affiliation{Department of Materials Science and Engineering, Cornell University, Ithaca, New York 14853, USA}
\affiliation{Kavli Institute at Cornell for Nanoscale Science, Cornell University, Ithaca, New York 14853, USA}
\begin{abstract}
Ultrawide bandgap heterojunction p-n diodes with polarization-induced AlGaN p-type layers are demonstrated using plasma-assisted molecular beam epitaxy on bulk AlN substrates. Current-voltage characteristics show a turn on voltage of $V_{\text{bi}}\approx5.5$~V, a minimum room temperature ideality factor of $\eta\approx 1.63$, and more than 12 orders of current modulation at room temperature.  Stable current operation of the ultrawide bandgap semiconductor diode is measured up to a temperature of 300$^\circ$C.  The one-sided n$^{+}$-p heterojunction diode design enables a direct measurement of the spatial distribution of polarization-induced mobile hole density in the graded AlGaN layer from the capacitance-voltage profile. The measured average mobile hole density is $p \sim 5.7 \times 10^{17}$~cm$^{-3}$, in close agreement with what is theoretically expected from distributed polarization doping. Light emission peaked at 260~nm (4.78~eV) observed in electroluminescence corresponds to interband radiative recombination in the n$^{+}$ AlGaN layer. A much weaker deep-level emission band observed at 3.4~eV is attributed to cation-vacancy and silicon complexes in the heavily Si-doped AlGaN layer. These results demonstrate that distributed polarization doping enables ultrawide bandgap semiconductor heterojunction p-n diodes that have wide applications ranging from power electronics to deep-ultraviolet photonics. These devices can operate at high temperatures and in harsh environments. 
\end{abstract}

\maketitle
Aluminium gallium nitride (AlGaN) based p-n junction diodes are promising devices for advancing high power electronics and deep ultraviolet (UV) photonics. These applications are enabled by the following desirable properties of the AlGaN semiconductor system: a large tunable direct energy bandgap (3.4-6.2~eV), high critical electric field (3-15~MV/cm), and high thermal conductivity (260~W/m.K for GaN and 340~W/m.K for AlN), amongst others. p-type doping is the major bottleneck in realising a p-n junction using these ultrawide bandgap semiconductors. The acceptor ionization energy of magnesium~(Mg) increases with the energy bandgap from GaN ($E_{a} \sim$200~meV\cite{brochenDependenceMgrelatedAcceptor2013,gotzActivationAcceptorsMg1996}) to AlN ($E_{a} \sim$630~meV \cite{taniyasuAluminiumNitrideLightemitting2006}). Recent reports on p-type doping of AlN with lower acceptor binding energies and new shallow dopants like Be are under further investigation\cite{ahmadSubstantialPTypeConductivity2021,ishiiRevisitingSubstitutionalMg2023}. 

Deep acceptors with $E_{a} >> k_{b}T$ lead to poor ionization, low free carrier concentrations, and increased on-resistance in the diode p-type region. The large disparity between electron and hole carrier concentration and mobility causes efficiency droop in light emitting diodes (LEDs) \cite{parkFundamentalLimitationsWideBandgap2018a}. Several allied limitations with using a high concentration of Mg in active regions include surface polarity inversion \cite{liSurfacePolarityDependence2000,greenPolarityControlMolecular2003}, Mg precipitation\cite{hansenObservationMgRichPrecipitates2001}, surface segregation\cite{xingMemoryEffectRedistribution2003}, self-compensating defects formation\cite{figgeMagnesiumSegregationFormation2002}, memory effects\cite{ohbaStudyStrongMemory1994}, and increased frequency dispersion in diodes \cite{kozodoyDepletionRegionEffects2000}, among others. In the design of waveguides for deep-UV laser diodes, Mg-doped cladding layers are undesirable because they cause high Mg-induced optical losses that increase the threshold gain \cite{martensEffectsMagnesiumDoping2017}.

One way to overcome these challenges is to leverage the spontaneous and piezoelectric polarization of wurtzite III/V nitride semiconductors. Spatial compositional grading of AlGaN along the polar axis creates a {\em fixed} bulk 3D polarization bound charge, whose electric field enables the formation of {\em mobile} 3-dimensional charges of opposite polarity. The use of such distributed polarization doped (DPD) layers has played a key role in the demonstration of deep UV laser diodes \cite{zhangKeyTemperaturedependentCharacteristics2022,zhangContinuouswaveLasingAlGaNbased2022,zhang271NmDeepultraviolet2019,tanakaLowthresholdcurrent85MA2021}, LEDs \cite{leeEfficientInGaNPContacts2019,kolbe234NmFarultravioletC2023} and power diodes\cite{kumabeSpaceChargeProfiles2022,nomotoDistributedPolarizationdopedGaN2022,huUnityIdealityFactor2015,liPolarizationInducedPnjunction2012}. In this work we create a DPD p-type layer by linearly grading down the AlGaN composition along the $+$c direction of the crystal. The grading creates a 3-dimensional hole gas\cite{simonPolarizationInducedHoleDoping2010,ahmadPolarizationDopingInitio2022,simonPolarizationInducedHoleDoping2010,chaudhuriPolarizationinduced2DHole2019}. We use a one-sided n$^{+}$-p heterojunction to measure the spatial density profile of the 3D hole gas. We also find that the resulting ultrawide bandgap heterojunction diode current-voltage characteristics exhibit close to unity ideality factor, stable high temperature operation, and electroluminescence.

\begin{figure*}
\includegraphics[width=\textwidth]{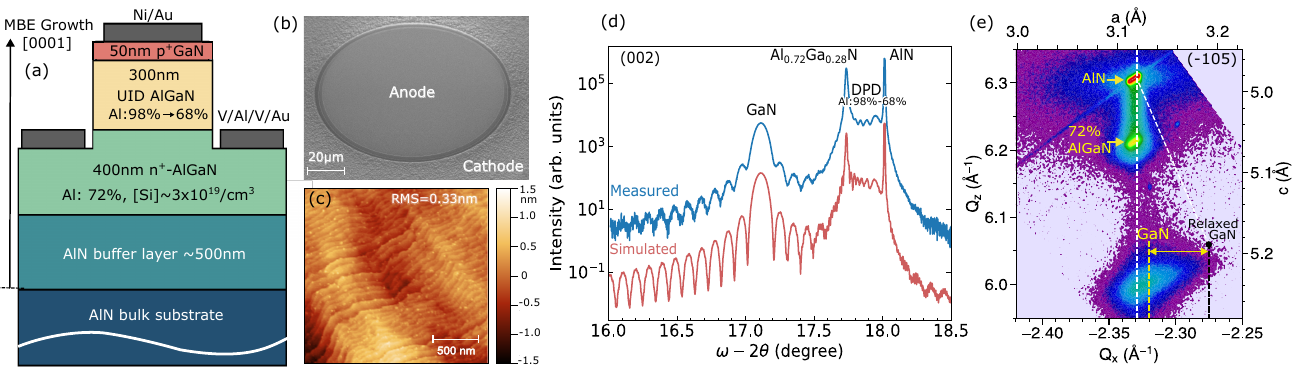}
\caption{\label{fig:1} (a) Cross-sectional view of the fabricated device structure in this study. (b) SEM image of a 104~\textmu m diameter device. (c) $2\times2$~\textmu m$^2$ AFM micrographs of the top GaN surface of the as-grown sample indicating smooth surface and 2 dimensional growth mode.  (d)~Measured and simulated $2\theta-\omega$ XRD scans of the sample across the (002) diffractions. (e) RSMs across the assymetric (-105) diffractions.}
\end{figure*}

Space-charge profiling of DPD-based diodes grown on bulk AlN was recently reported in metal organic chemical vapor deposition (MOCVD) grown diodes\cite{zhangSpaceChargeProfile2020}. There are no reports of such measurements on molecular beam epitaxial (MBE) grown devices. MBE offers some differences from MOCVD such as lower growth temperature, lower hydrogen incorporation, and the absence of memory effects enabling a precise dopant profiles and sharp heterointerfaces. While MBE grown 2D hole gases were demonstrated on single crystal AlN bulk substrates \cite{zhangPolarizationinduced2DHole2021}, polarization induced 3D hole gases on bulk AlN have not been realized yet. In this study we present an MBE grown quasi-vertical p-n diode that uses an undoped distributed polarization doped layer for hole injection. We find an average mobile hole concentration of 5.7$\times$10$^{17}$~cm$^{-3}$, consistent with what is expected from spontaneous and piezoelectric polarization effects. These findings make unintentionally doped DPD based diodes an attractive alternative to the conventional impurity based pn diodes.

The diode heterostructures were grown in a nitrogen plasma-assisted Veeco Gen10 molecular beam epitaxy (MBE) system on $+$c-plane single crystal bulk AlN substrates. The substrates were subjected to two essential cleaning steps as described in detail elsewhere \cite{choMolecularBeamHomoepitaxy2020,leeSurfaceControlMBE2020}: (1) an ex-situ cleaning using solvents and acids, and (2) an in-situ cleaning achieved through repeated cycles of Al adsorption and desorption, referred to as Al-assisted polishing. These steps eliminate the native surface oxides to enable high-quality homoepitaxy.

As shown in Fig.~\ref{fig:1}~(a), a 500~nm thick AlN buffer layer was grown at a high temperature of  $\mathrm{T_{sub}}$$\sim$1060~$^\circ$C in Al-rich conditions to isolate the device layers from remaining substrate surface impurities. The subsequent AlGaN epilayer was grown under Ga-rich conditions at a lower substrate temperature of 880~$^\circ$C to enhance Ga incorporation. Excess metal was thermally desorbed at the end of each layer to ensure sharp heterojunctions. From bottom to top along the metal-polar growth direction, the targeted p-n diode heterostructure as seen in Fig.~\ref{fig:1}~(a) consists of: (1) a 500~nm MBE grown AlN buffer layer, (2) a 400~nm Al$_{0.7}$Ga$_{0.3}$N layer with Si doping density $N_{d} \approx 3\times10^{19}$~cm$^{-3}$ which resulted in a free electron density $n \approx $ $2\times10^{19}$~cm$^{-3}$ at room temperature from Hall-effect measurements, (3) an  unintentionally doped (UID) DPD layer linearly graded from Al$_{0.95}$Ga$_{0.05}$N to Al$_{0.65}$Ga$_{0.35}$N over 300~nm, followed by (4) a 50 nm heavily Mg-doped GaN capping layer to form the metal p-contacts. In the entire device stack, the Mg impurity doping is only incorporated in the GaN p-contact layer and not in the Al containing UWBG layers. 

\begin{figure*}
\includegraphics[width=0.75\textwidth]{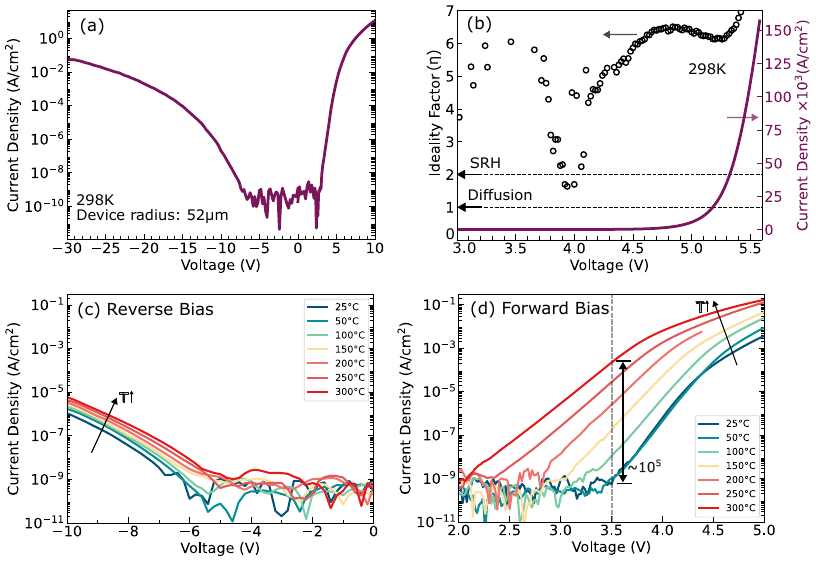}
\caption{\label{fig:2} (a) Room temperature J-V characteristics of the diode. (b) The extracted ideality factor from (a) using equation \eqref{eq:1}. (c) Temperature dependent J-V characteristics in reverse bias. (d) Temperature dependent J-V characteristics in forward bias.}
\end{figure*}

Following epitaxy of the device heterostructures, the layers were fabricated into quasi-vertical diodes as indicated in Fig.~\ref{fig:1}~(a) and shown in Fig.~\ref{fig:1}~(b). First, circular device mesas were formed by chlorine based inductively coupled plasma reactive ion etching (ICP-RIE) with a total etch depth extending 100~nm into the n-type AlGaN layer. The device mesa diameters range from 20 to 400~\textmu m. Then, n-type metal-semiconductor contacts were formed by electron beam evaporation of a V/Al/V/Au stack of 20/80/40/100~nm thickness, which was subsequently rapid thermal annealed (RTA) at 800~$^{\circ}$C for 60 seconds in a N$_{2}$ ambience. Finally, p-type metal-semiconductor Ni/Au contacts of 15/20~nm thickness were deposited by electron beam evaporation and annealed by RTA at 450~$^{\circ}$C for 30 seconds in an O$_{2}$ environment. A 20/100~nm Ti/Au metal stack was subsequently deposited by electron beam evaporation for probing. Electrical measurements were performed using a Keithley 4200A semiconductor parameter analyzer. Electroluminescence spectra were collected using a Princeton Instruments spectrometer with 2400~grooves/mm and a blaze wavelength of 240~nm. All measurements were performed on devices with a 104~\textmu m diameter, unless specified otherwise. Figure~\ref{fig:1}~(a) shows a cross-sectional view of the device, and Fig.~\ref{fig:1}~(b) shows a scanning electron microsope (SEM) image of the fabricated device at a 45$^{\circ}$ tilt. The SEM was taken by a Zeiss ULTRA microscope at 5~kV beam voltage with an in-lens detector.

Fig.~\ref{fig:1}~(c) shows atomic steps on the top p-GaN layer of the as-grown sample with a root mean square roughness of 0.33~nm over $2\times2$~\textmu m$^2$ scan area measured by atomic force microscopy (AFM), confirming step flow growth mode throughout the structure. The X-ray diffraction (XRD) scans in Figure~\ref{fig:1}~(d), performed using a Panalytical Empyrean system, shows good agreement between the measured and simulated $2\theta-\omega$ XRD scans of the sample across the (002) diffractions. The actual Al compositions found from XRD are 2-3\% higher than the targeted structure. The reciprocal space map (RSM) around the asymmetric ($\bar{1}05$) diffractions in Fig.~\ref{fig:1}~(e) shows that the AlGaN layers are fully strained to the AlN, while the GaN layer is relaxed and has an in-plane lattice strain of approximately $1.9\%$.  


Fig.~\ref{fig:2}(a) shows the room temperature current-voltage characteristics of the diode. The reverse bias leakage current detection is limited by the 100~fA noise floor of the equipment until approximately -8~V, beyond which it increases gradually. In the forward bias, the diode turn on voltage is approximately 5.5~V, and a specific on-resistance of 0.9~$\Omega \cdot $cm$^2$ at 6~V. The maximum measured forward current density of 1.3~kA/cm$^2$ at $\sim 20$~V was limited by the current limit of the equipment. The measured 12 orders of current modulation (limited by the measurement noise floor and compliance) illustrates the capability of the AlGaN heterojunction p-n diode without Mg doping in the AlGaN layers.

Fig.~\ref{fig:2}~(b) shows the turn-on behavior in linear scale, and the corresponding ideality factor. The diode forward current is \cite{szePhysicsSemiconductorDevices2021} $J_{F} \approx J_0 \exp{[qV/(\eta k_b T)]}$, where $J_0$ is a voltage-independent and material-dependent coefficient, $q$ is the electron charge, $V$ is the junction voltage, and $\eta$ is the ideality factor. $\eta = 2$  when non-radiative Shockley-Read-Hall interband recombination current is dominant, and $\eta = 1$ when minority carrier diffusion current dominates. The voltage dependent ideality factor from the general diode relation
\begin{equation}\label{eq:1}
    \eta = \dfrac{q}{k_b T}\times\dfrac{dV}{d\ln{(J/J_0)}}
\end{equation} 
is used to obtain the $\eta$ shown in Fig.~\ref{fig:2}~(b). $\eta$ reaches a minimum value of approximately 1.63 in a narrow voltage range around 4~V near turn on, one of the lowest reported to date in ultrawide bandgap pn diodes. The deviation from the experimental ideality factor from the theoretical models ($\mathrm{1\le\eta\le2}$) in AlGaN/GaN p-n junction diodes has been attributed to non-ohmic metal-semiconductor junctions\cite{kumabeSpaceChargeProfiles2022,shahExperimentalAnalysisTheoretical2003}. By performing transfer length method (TLM) measurements we found that both the p and n contacts are not entirely ohmic, and exhibit some non-linearity. In this case the total ideality factor is the sum of individual ideality factors of all the rectifying junctions in the system as derived by Shah et al\cite{shahExperimentalAnalysisTheoretical2003}. The presence of the Schottky-like contact diodes in series with the pn junction diode therefore complicates an accurate determination of the true ideality factor of the pn junction itself.

Figs.~\ref{fig:2}~(c) and (d) show the temperature-dependence of the diode current from 25~$^\circ$C to 300~$^\circ$C. The reverse leakage current in Fig.~\ref{fig:2}~(c) increases with increasing electric field and temperature. This is a signature of trap-assisted tunneling being the dominant leakage mechanisms, such as the Frenkel-Poole (FP) process \cite{ferdousEffectThreadingDefects2007} or variable-range hopping (VRH)\cite{mottConductionNoncrystallineMaterials1969}. A far more dramatic temperature dependence is seen in the forward bias current in Fig.~\ref{fig:2}~(d). For example, at 3.5~V forward bias the current density increases by 5 orders of magnitude when the temperature is increased from 25~$^\circ$C to 300~$^\circ$C. A large exponential increase in current is indeed expected with temperature, because the intrinsic interband thermally generated carrier density is $n_i \propto \exp{ (- E_g / (2k_b T))}$, and in the ideal diode theory $J_0 \propto n_{i}^2 \propto \exp{ (- E_g / (k_b T)) }$ is a strong function of temperature. 

\begin{figure*}
\includegraphics[width=\textwidth]{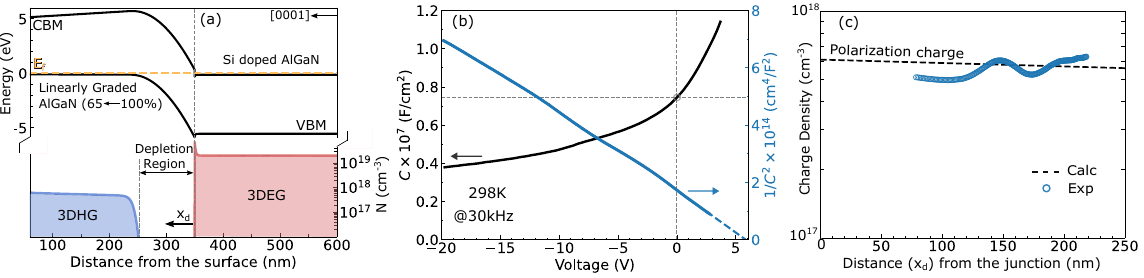}
\caption{\label{fig:3} (a) Energy band diagram and free electron and hole concentration of the p-n diode at zero bias. (b) Room-temperature capacitance-voltage measurements at 30~kHz AC frequency. (c) The extracted charge-density profile in the graded AlGaN layer using equation \eqref{eq:5} which matches well with the polarization charge calculations done using equations \eqref{eq:3}-\eqref{eq:4}.}
\end{figure*}

Fig.~\ref{fig:3}~(a) shows the calculated energy band diagram of the pn heterojunction diode at zero bias, highlighting the depletion region. Unlike in non-polar pn diodes, at this polar AlN/AlGaN p-n heterojunction there is no depletion region in the n-side.  Across the heterointerface of n-Al$_{0.72}$Ga$_{0.28}$N and AlN, there is a polarization discontinuity {\em and} an energy band discontinuity.  Since the n-AlGaN is doped with donors, the combination gives rise to a 2-dimensional electron gas (2DEG) of density $\sim$1.64$\times$10$^{13}$~cm$^{-2}$ at the heterojunction. Because of this n-type {\em accumulation} region, the depletion region falls completely in the p-side. Furthermore, the mobile holes in the linearly graded AlGaN layer are due to distributed polarization doping. Thus, capacitance-voltage (CV) profiling should unambiguously extract the charge-density profile in the DPD layer. The low reverse bias leakage in this device (Fig.~\ref{fig:2}~(a))  enables reliable CV measurements up to -20~V. The dc bias determines the depletion depth, and a 30~mV AC signal at a frequency of 30~kHz was used for the capacitance measurement in a standard parallel capacitance and conductance $C_p - G_p$ model\cite{staufferFundamentalsSemiconductorCV2008}. Figure~\ref{fig:3}~(b) shows the measured capacitance as a function of the applied DC bias at room temperature. The loss tangent $\tan \delta = 2 \pi f (C_p / G_p)$ remains below 0.1 in the entire voltage range, ensuring the validity of the data. The built-in voltage of the junction from the extrapolation of $1/C^{2}$ vs $V$ in Fig.~\ref{fig:3}~(b) is 5.8~V, close to the expected value of 5.5~V. 

The 3D bulk polarization charge density in the DPD layer is the sum of both spontaneous and piezoelectric polarizations $P_{tot}=P_{PZ}+P_{SP}$. The piezoelectric polarization of Al$_x$Ga$_{1-x}$N coherently strained on AlN is \cite{woodPolarizationEffectsSemiconductors2007},
\begin{equation}\label{eq:3}
    P_{PZ}(x) = 2\times\Big(\dfrac{a_{\mathrm{Al_xGa_{1-x}N}}-a_{\mathrm{AlN}}}{a_{\mathrm{AlN}}}\Big)\times \Big(e_{31}-e_{33}\dfrac{c_{13}}{c_{33}}\Big),
\end{equation}
where $c_{13}$ and $c_{33}$ are elastic coefficients and $e_{31}$ and $e_{33}$ are piezoelectric moduli. The values of spontaneous polarization, elastic coefficients and piezoelectric moduli for AlN and GaN were taken from Table~1 of Ref.\citenum{zhangSpaceChargeProfile2020}. The corresponding values for Al$_{x}$Ga$_{1-x}$N were obtained by linear interpolation (Vegard's law). The net carrier-density profile in cm$^{-3}$ along the [0001] direction (z~axis) is,
\begin{equation}\label{eq:4}
    \rho(z)=\dfrac{1}{q} \nabla \cdot P_{tot}=\dfrac{1}{q}\dfrac{\partial P(x(z))}{\partial z} ,
\end{equation}
where $x(z)$ is the graded Al-content profile along the z~axis, a linear function in this case. The charge-density at the edge of the depletion region is extracted from the measured CV data of a one-sided abrupt junction \cite{szePhysicsSemiconductorDevices2021}
\begin{equation}\label{eq:5}
    N= \dfrac{-2}{q\epsilon_s\epsilon_0}\times\Bigg[\dfrac{1}{d(1/C^2)/dV}\Bigg],
\end{equation}
where $q$ is the electron charge, $\epsilon_{s}$ is the relative permittivity of the semiconductor at the edge of the depletion region, and $\epsilon_0$ is the permittivity of vacuum. A constant value of 9.35 was used for $\epsilon_{s}$ corresponding to an average Al composition of 83\% in the DPD layer, interpolated between AlN ($\epsilon_{s}$=9.21 \cite{fenebergAnisotropicAbsorptionEmission2013}) and GaN ($\epsilon_{s}$=10.04 \cite{fenebergBandGapRenormalization2014}). The depletion width in the DPD layer is $W_{D}=(\epsilon_0 \epsilon_s)/C$.

Fig.~\ref{fig:3}~(b) shows the experimentally measured, and the calculated charge-density profile (dashed line) along the $z$ direction. The experimental average charge density $5.7\times10^{17}$~cm$^{-3}$ is approximately equal to the calculated density of $5.8\times10^{17}$~cm$^{-3}$. Thus the presence of a high density polarization induced 3D hole gas close to the theoretically predicted density is observed. The rather interesting oscillations of the charge density observed in all devices are not captured in the simulation. They could originate either due to periodic fluctuations in Al composition, or due Friedel oscillations of the three-dimensional hole gas\cite{kanisawaImagingFriedelOscillation2001,vanderwielenDirectObservationFriedel1996}. The root of these oscillations will be investigated in a future work.

\begin{figure*}
\includegraphics[width=0.8\textwidth]{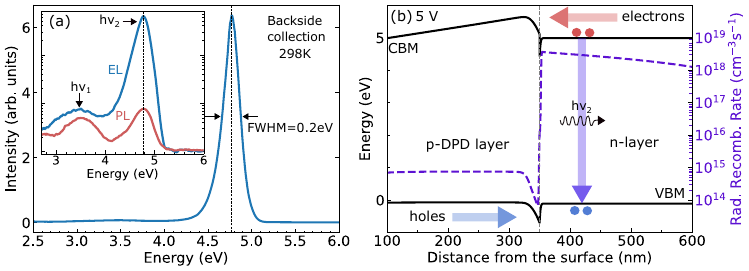}
\caption{\label{fig:4} (a) Room temperature electroluminescence collected from the backside of the device. Inset shows the comparison (in logarithmic scale) between EL from the device and PL from an AlGaN sample with the same composition and doping as the n-layer of the device. (b)~Energy band diagram of the diode at 5~V forward bias with the arrow showing the radiative transition corresponding to the main emission peak.}
\end{figure*}

Fig.~\ref{fig:4}~(a) shows the measured room temperature electroluminescence (EL) collected from the backside of large \SI{400}{\micro\meter} diameter devices at a forward current density of 110 A/cm$^2$ at room temperature. A peak at 4.78 eV dominates the emission spectrum. Additionally, a far less intense deep-level luminescence peak of energy $\sim$3.4 eV is also observed. To identify the origins of these peaks, room temperature photoluminescence (PL) experiments were also conducted on an Al$_{0.72}$Ga$_{0.28}$N/AlN sample with the same Si doping density and without the DPD and p-contact layers using a 193 nm ArF excimer laser excitation. The insert in Fig.~\ref{fig:4}~(a) shows a comparison of the EL and PL spectra. It confirms that the dominant emission peak in EL is from interband radiative recombination in the Si-doped Al$_{0.72}$Ga$_{0.28}$N layer. 

Fig.~\ref{fig:4}~(b) shows the calculated energy-band diagram of the diode at a junction bias of 5~V, along with the spatially resolved radiative recombination rate, simulated using STR SiLENSe \cite{SiLENSe}. The purple arrow on the plot indicates the interband transition responsible for the dominant peak in the EL spectra, where the radiative recombination rate is nearly $10^3$ times more intense than in the p-DPD layer. The low to non-observable emission from the DPD layer in the EL spectrum is due to two reasons: (1) the product of electron and hole concentrations under forward bias is significantly higher in the n-layer leading to a higher radiative recombination rate since $ R \propto np$, and (2) the luminescence resulting from recombination within the DPD layer has higher energy than the energy bandgap of Al$_{0.72}$Ga$_{0.28}$N.  This means photons emitted in the p-DPD layer moving towards the bulk will be absorbed and re-emitted at a photon energy equal to the n-layer energy bandgap during backside collection. 

The weak sub-bandgap peak at approximately 3.4~eV is very close to the energy bandgap of GaN. This peak could be due to optical excitation of the top GaN layer from the emitted 4.8~eV photons which then make it across the wafer to the backside collector. But the appearance of a weak 3.4 eV peak in the inset of Fig.~\ref{fig:4}~(b) in the PL spectra of Al$_{0.72}$Ga$_{0.28}$N without any GaN layer indicates the EL peak is also from the n-AlGaN layer. Chichibu et al.\cite{chichibuImpactsSidopingResultant2013} have proposed the existence of defect complexes consisting of cation vacancies and silicon (V$_{\text{III}}$-nSi$_{\text{III}}$) as an explanation for the deep PL emission bands. These complexes act as self-compensating acceptor-type defects in Si doped AlN and AlGaN \cite{hyunkimTrackingPointDefects2023,prozheevElectricalCompensationCation2020}. But it should be noted that literature reports on luminescence in AlN and AlGaN grown by MBE are scarce \cite{vandeurzenExcitonicDeeplevelEmission2023,kanekoOpticalPropertiesHighly2013,vandeurzenOpticallyPumpedDeepUV2022,jmerikPlasmaassistedMolecularBeam2013}, and that defect formation may strongly depend on the method of deposition.

In summary, ultrawide bandgap semiconductor diodes exhibiting low reverse bias leakage and high on/off ratio are realized by MBE, thanks to the low dislocation-density of the epilayers grown on bulk AlN substrates. Completely one-sided p-n heterojunction diodes are realized exploiting polarization induced doping on the n-side to remove the depletion layer, and distributed polarization doping instead of Mg acceptor doping for the p-type depletion layer. Through capacitance-voltage measurements, the mobile hole concentration and their spatial distribution in the graded AlGaN layers was directly measured and is found to be consistent with what is expected from polarization effects. These polarization-induced ultrawide bandgap semiconductor diodes show stable performance up to 300~$^\circ$C. The electroluminescence from these diodes is dominated by interband radiative recombination, and deep level luminescence is greatly suppressed. This suggests the presence of low point defect densities in the MBE-grown Si-doped AlGaN layer. Overall this study demonstrates the flexibility in the design of new kinds of p-n heterojunction diodes through polarization-induced doping to achieve properties that are not possible in standard diodes. Such heterostructure design that combine bandgap engineering intimately with polarization engineering opens opportunities for more efficient photonic and electronic devices with ultrawide bandgap polar semiconductors than what is possible in nonpolar semiconductors.
\section*{acknowledgments}
The authors thank Takeru Kumabe from Nagoya University for helpful discussions. This work was supported as part of the Ultra Materials for
a Resilient Energy Grid, an Energy Frontier Research Center funded by the U.S. Department of Energy, Office of Science, Basic Energy Sciences under Award \#DE-SC0021230. The authors acknowledge the use of the Cornell NanoScale Facility (CNF) a member of the National Nanotechnology Co-ordinated Infrastructure (NNCI), which is supported by the National Science Foundation (NSF Grant NNCI-2025233). 
\section*{author declarations}
\subsection*{Conflict of Interest}
The authors have no conflict of interest to declare.
\section*{Data Availability Statement}
The data that support the findings of this study are available from the corresponding authors upon reasonable request.

\nocite{*}
\bibliography{DPDDiodefinal}

\begin{thebibliography}{49}%
\makeatletter
\providecommand \@ifxundefined [1]{%
 \@ifx{#1\undefined}
}%
\providecommand \@ifnum [1]{%
 \ifnum #1\expandafter \@firstoftwo
 \else \expandafter \@secondoftwo
 \fi
}%
\providecommand \@ifx [1]{%
 \ifx #1\expandafter \@firstoftwo
 \else \expandafter \@secondoftwo
 \fi
}%
\providecommand \natexlab [1]{#1}%
\providecommand \enquote  [1]{``#1''}%
\providecommand \bibnamefont  [1]{#1}%
\providecommand \bibfnamefont [1]{#1}%
\providecommand \citenamefont [1]{#1}%
\providecommand \href@noop [0]{\@secondoftwo}%
\providecommand \href [0]{\begingroup \@sanitize@url \@href}%
\providecommand \@href[1]{\@@startlink{#1}\@@href}%
\providecommand \@@href[1]{\endgroup#1\@@endlink}%
\providecommand \@sanitize@url [0]{\catcode `\\12\catcode `\$12\catcode
  `\&12\catcode `\#12\catcode `\^12\catcode `\_12\catcode `\%12\relax}%
\providecommand \@@startlink[1]{}%
\providecommand \@@endlink[0]{}%
\providecommand \url  [0]{\begingroup\@sanitize@url \@url }%
\providecommand \@url [1]{\endgroup\@href {#1}{\urlprefix }}%
\providecommand \urlprefix  [0]{URL }%
\providecommand \Eprint [0]{\href }%
\providecommand \doibase [0]{https://doi.org/}%
\providecommand \selectlanguage [0]{\@gobble}%
\providecommand \bibinfo  [0]{\@secondoftwo}%
\providecommand \bibfield  [0]{\@secondoftwo}%
\providecommand \translation [1]{[#1]}%
\providecommand \BibitemOpen [0]{}%
\providecommand \bibitemStop [0]{}%
\providecommand \bibitemNoStop [0]{.\EOS\space}%
\providecommand \EOS [0]{\spacefactor3000\relax}%
\providecommand \BibitemShut  [1]{\csname bibitem#1\endcsname}%
\let\auto@bib@innerbib\@empty
\bibitem [{\citenamefont {Brochen}\ \emph {et~al.}(2013)\citenamefont
  {Brochen}, \citenamefont {Brault}, \citenamefont {Chenot}, \citenamefont
  {Dussaigne}, \citenamefont {Leroux},\ and\ \citenamefont
  {Damilano}}]{brochenDependenceMgrelatedAcceptor2013}%
  \BibitemOpen
  \bibfield  {author} {\bibinfo {author} {\bibfnamefont {S.}~\bibnamefont
  {Brochen}}, \bibinfo {author} {\bibfnamefont {J.}~\bibnamefont {Brault}},
  \bibinfo {author} {\bibfnamefont {S.}~\bibnamefont {Chenot}}, \bibinfo
  {author} {\bibfnamefont {A.}~\bibnamefont {Dussaigne}}, \bibinfo {author}
  {\bibfnamefont {M.}~\bibnamefont {Leroux}},\ and\ \bibinfo {author}
  {\bibfnamefont {B.}~\bibnamefont {Damilano}},\ }\bibfield  {title} {\enquote
  {\bibinfo {title} {Dependence of the {{Mg-related}} acceptor ionization
  energy with the acceptor concentration in p-type {{GaN}} layers grown by
  molecular beam epitaxy},}\ }\href {https://doi.org/10.1063/1.4813598}
  {\bibfield  {journal} {\bibinfo  {journal} {Appl. Phys. Lett.}\ }\textbf
  {\bibinfo {volume} {103}},\ \bibinfo {pages} {032102} (\bibinfo {year}
  {2013})}\BibitemShut {NoStop}%
\bibitem [{\citenamefont {G{\"o}tz}\ \emph {et~al.}(1996)\citenamefont
  {G{\"o}tz}, \citenamefont {Johnson}, \citenamefont {Walker}, \citenamefont
  {Bour},\ and\ \citenamefont {Street}}]{gotzActivationAcceptorsMg1996}%
  \BibitemOpen
  \bibfield  {author} {\bibinfo {author} {\bibfnamefont {W.}~\bibnamefont
  {G{\"o}tz}}, \bibinfo {author} {\bibfnamefont {N.~M.}\ \bibnamefont
  {Johnson}}, \bibinfo {author} {\bibfnamefont {J.}~\bibnamefont {Walker}},
  \bibinfo {author} {\bibfnamefont {D.~P.}\ \bibnamefont {Bour}},\ and\
  \bibinfo {author} {\bibfnamefont {R.~A.}\ \bibnamefont {Street}},\ }\bibfield
   {title} {\enquote {\bibinfo {title} {Activation of acceptors in {{Mg}}-doped
  {{GaN}} grown by metalorganic chemical vapor deposition},}\ }\href
  {https://doi.org/10.1063/1.116503} {\bibfield  {journal} {\bibinfo  {journal}
  {Appl. Phys. Lett.}\ }\textbf {\bibinfo {volume} {68}},\ \bibinfo {pages}
  {667--669} (\bibinfo {year} {1996})}\BibitemShut {NoStop}%
\bibitem [{\citenamefont {Taniyasu}, \citenamefont {Kasu},\ and\ \citenamefont
  {Makimoto}(2006)}]{taniyasuAluminiumNitrideLightemitting2006}%
  \BibitemOpen
  \bibfield  {author} {\bibinfo {author} {\bibfnamefont {Y.}~\bibnamefont
  {Taniyasu}}, \bibinfo {author} {\bibfnamefont {M.}~\bibnamefont {Kasu}},\
  and\ \bibinfo {author} {\bibfnamefont {T.}~\bibnamefont {Makimoto}},\
  }\bibfield  {title} {\enquote {\bibinfo {title} {An aluminium nitride
  light-emitting diode with a wavelength of 210 nanometres},}\ }\href
  {https://doi.org/10.1038/nature04760} {\bibfield  {journal} {\bibinfo
  {journal} {Nature}\ }\textbf {\bibinfo {volume} {441}},\ \bibinfo {pages}
  {325--328} (\bibinfo {year} {2006})}\BibitemShut {NoStop}%
\bibitem [{\citenamefont {Ahmad}\ \emph {et~al.}(2021)\citenamefont {Ahmad},
  \citenamefont {Lindemuth}, \citenamefont {Engel}, \citenamefont {Matthews},
  \citenamefont {McCrone},\ and\ \citenamefont
  {Doolittle}}]{ahmadSubstantialPTypeConductivity2021}%
  \BibitemOpen
  \bibfield  {author} {\bibinfo {author} {\bibfnamefont {H.}~\bibnamefont
  {Ahmad}}, \bibinfo {author} {\bibfnamefont {J.}~\bibnamefont {Lindemuth}},
  \bibinfo {author} {\bibfnamefont {Z.}~\bibnamefont {Engel}}, \bibinfo
  {author} {\bibfnamefont {C.~M.}\ \bibnamefont {Matthews}}, \bibinfo {author}
  {\bibfnamefont {T.~M.}\ \bibnamefont {McCrone}},\ and\ \bibinfo {author}
  {\bibfnamefont {W.~A.}\ \bibnamefont {Doolittle}},\ }\bibfield  {title}
  {\enquote {\bibinfo {title} {Substantial {{P-Type Conductivity}} of {{AlN
  Achieved}} via {{Beryllium Doping}}},}\ }\href
  {https://doi.org/10.1002/adma.202104497} {\bibfield  {journal} {\bibinfo
  {journal} {Adv. Mater.}\ }\textbf {\bibinfo {volume} {33}},\ \bibinfo {pages}
  {2104497} (\bibinfo {year} {2021})}\BibitemShut {NoStop}%
\bibitem [{\citenamefont {Ishii}\ \emph {et~al.}(2023)\citenamefont {Ishii},
  \citenamefont {Yoshikawa}, \citenamefont {Funato},\ and\ \citenamefont
  {Kawakami}}]{ishiiRevisitingSubstitutionalMg2023}%
  \BibitemOpen
  \bibfield  {author} {\bibinfo {author} {\bibfnamefont {R.}~\bibnamefont
  {Ishii}}, \bibinfo {author} {\bibfnamefont {A.}~\bibnamefont {Yoshikawa}},
  \bibinfo {author} {\bibfnamefont {M.}~\bibnamefont {Funato}},\ and\ \bibinfo
  {author} {\bibfnamefont {Y.}~\bibnamefont {Kawakami}},\ }\bibfield  {title}
  {\enquote {\bibinfo {title} {Revisiting the substitutional {{Mg}} acceptor
  binding energy of {{AlN}}},}\ }\href
  {https://doi.org/10.1103/PhysRevB.108.035205} {\bibfield  {journal} {\bibinfo
   {journal} {Phys. Rev. B}\ }\textbf {\bibinfo {volume} {108}},\ \bibinfo
  {pages} {035205} (\bibinfo {year} {2023})}\BibitemShut {NoStop}%
\bibitem [{\citenamefont {Park}\ \emph {et~al.}(2018)\citenamefont {Park},
  \citenamefont {Kim}, \citenamefont {Schubert}, \citenamefont {Cho},\ and\
  \citenamefont {Kim}}]{parkFundamentalLimitationsWideBandgap2018a}%
  \BibitemOpen
  \bibfield  {author} {\bibinfo {author} {\bibfnamefont {J.~H.}\ \bibnamefont
  {Park}}, \bibinfo {author} {\bibfnamefont {D.~Y.}\ \bibnamefont {Kim}},
  \bibinfo {author} {\bibfnamefont {E.~F.}\ \bibnamefont {Schubert}}, \bibinfo
  {author} {\bibfnamefont {J.}~\bibnamefont {Cho}},\ and\ \bibinfo {author}
  {\bibfnamefont {J.~K.}\ \bibnamefont {Kim}},\ }\bibfield  {title} {\enquote
  {\bibinfo {title} {Fundamental {{Limitations}} of {{Wide-Bandgap
  Semiconductors}} for {{Light-Emitting Diodes}}},}\ }\href
  {https://doi.org/10.1021/acsenergylett.8b00002} {\bibfield  {journal}
  {\bibinfo  {journal} {ACS Energy Lett.}\ }\textbf {\bibinfo {volume} {3}},\
  \bibinfo {pages} {655--662} (\bibinfo {year} {2018})}\BibitemShut {NoStop}%
\bibitem [{\citenamefont {Li}\ \emph {et~al.}(2000)\citenamefont {Li},
  \citenamefont {Jurkovic}, \citenamefont {Wang}, \citenamefont {Van~Hove},\
  and\ \citenamefont {Chow}}]{liSurfacePolarityDependence2000}%
  \BibitemOpen
  \bibfield  {author} {\bibinfo {author} {\bibfnamefont {L.~K.}\ \bibnamefont
  {Li}}, \bibinfo {author} {\bibfnamefont {M.~J.}\ \bibnamefont {Jurkovic}},
  \bibinfo {author} {\bibfnamefont {W.~I.}\ \bibnamefont {Wang}}, \bibinfo
  {author} {\bibfnamefont {J.~M.}\ \bibnamefont {Van~Hove}},\ and\ \bibinfo
  {author} {\bibfnamefont {P.~P.}\ \bibnamefont {Chow}},\ }\bibfield  {title}
  {\enquote {\bibinfo {title} {Surface polarity dependence of {{Mg}} doping in
  {{GaN}} grown by molecular-beam epitaxy},}\ }\href
  {https://doi.org/10.1063/1.126152} {\bibfield  {journal} {\bibinfo  {journal}
  {Appl. Phys. Lett.}\ }\textbf {\bibinfo {volume} {76}},\ \bibinfo {pages}
  {1740--1742} (\bibinfo {year} {2000})}\BibitemShut {NoStop}%
\bibitem [{\citenamefont {Green}\ \emph {et~al.}(2003)\citenamefont {Green},
  \citenamefont {Haus}, \citenamefont {Wu}, \citenamefont {Chen}, \citenamefont
  {Mishra},\ and\ \citenamefont {Speck}}]{greenPolarityControlMolecular2003}%
  \BibitemOpen
  \bibfield  {author} {\bibinfo {author} {\bibfnamefont {D.~S.}\ \bibnamefont
  {Green}}, \bibinfo {author} {\bibfnamefont {E.}~\bibnamefont {Haus}},
  \bibinfo {author} {\bibfnamefont {F.}~\bibnamefont {Wu}}, \bibinfo {author}
  {\bibfnamefont {L.}~\bibnamefont {Chen}}, \bibinfo {author} {\bibfnamefont
  {U.~K.}\ \bibnamefont {Mishra}},\ and\ \bibinfo {author} {\bibfnamefont
  {J.~S.}\ \bibnamefont {Speck}},\ }\bibfield  {title} {\enquote {\bibinfo
  {title} {Polarity control during molecular beam epitaxy growth of {{Mg-doped
  GaN}}},}\ }\href {https://doi.org/10.1116/1.1589511} {\bibfield  {journal}
  {\bibinfo  {journal} {J. Vac. Sci. Technol. B}\ }\textbf {\bibinfo {volume}
  {21}},\ \bibinfo {pages} {1804--1811} (\bibinfo {year} {2003})}\BibitemShut
  {NoStop}%
\bibitem [{\citenamefont {Hansen}\ \emph {et~al.}(2001)\citenamefont {Hansen},
  \citenamefont {Chen}, \citenamefont {Speck},\ and\ \citenamefont
  {DenBaars}}]{hansenObservationMgRichPrecipitates2001}%
  \BibitemOpen
  \bibfield  {author} {\bibinfo {author} {\bibfnamefont {M.}~\bibnamefont
  {Hansen}}, \bibinfo {author} {\bibfnamefont {L.}~\bibnamefont {Chen}},
  \bibinfo {author} {\bibfnamefont {J.}~\bibnamefont {Speck}},\ and\ \bibinfo
  {author} {\bibfnamefont {S.}~\bibnamefont {DenBaars}},\ }\bibfield  {title}
  {\enquote {\bibinfo {title} {Observation of {{Mg-Rich Precipitates}} in the
  p-{{Type Doping}} of {{GaN-Based Laser Diodes}}},}\ }\href
  {https://doi.org/10.1002/1521-3951(200111)228:2<353::AID-PSSB353>3.0.CO;2-Q}
  {\bibfield  {journal} {\bibinfo  {journal} {Phys. Status Solidi B}\ }\textbf
  {\bibinfo {volume} {228}},\ \bibinfo {pages} {353--356} (\bibinfo {year}
  {2001})}\BibitemShut {NoStop}%
\bibitem [{\citenamefont {Xing}\ \emph {et~al.}(2003)\citenamefont {Xing},
  \citenamefont {Green}, \citenamefont {Yu}, \citenamefont {Mates},
  \citenamefont {Kozodoy}, \citenamefont {Keller}, \citenamefont {DenBaars},\
  and\ \citenamefont {Mishra}}]{xingMemoryEffectRedistribution2003}%
  \BibitemOpen
  \bibfield  {author} {\bibinfo {author} {\bibfnamefont {H.}~\bibnamefont
  {Xing}}, \bibinfo {author} {\bibfnamefont {D.~S.}\ \bibnamefont {Green}},
  \bibinfo {author} {\bibfnamefont {H.}~\bibnamefont {Yu}}, \bibinfo {author}
  {\bibfnamefont {T.}~\bibnamefont {Mates}}, \bibinfo {author} {\bibfnamefont
  {P.}~\bibnamefont {Kozodoy}}, \bibinfo {author} {\bibfnamefont
  {S.}~\bibnamefont {Keller}}, \bibinfo {author} {\bibfnamefont {S.~P.}\
  \bibnamefont {DenBaars}},\ and\ \bibinfo {author} {\bibfnamefont {U.~K.}\
  \bibnamefont {Mishra}},\ }\bibfield  {title} {\enquote {\bibinfo {title}
  {Memory {{Effect}} and {{Redistribution}} of {{Mg}} into {{Sequentially
  Regrown GaN Layer}} by {{Metalorganic Chemical Vapor Deposition}}},}\ }\href
  {https://doi.org/10.1143/JJAP.42.50} {\bibfield  {journal} {\bibinfo
  {journal} {Jpn. J. Appl. Phys.}\ }\textbf {\bibinfo {volume} {42}},\ \bibinfo
  {pages} {50} (\bibinfo {year} {2003})}\BibitemShut {NoStop}%
\bibitem [{\citenamefont {Figge}\ \emph {et~al.}(2002)\citenamefont {Figge},
  \citenamefont {Kr{\"o}ger}, \citenamefont {B{\"o}ttcher}, \citenamefont
  {Ryder},\ and\ \citenamefont
  {Hommel}}]{figgeMagnesiumSegregationFormation2002}%
  \BibitemOpen
  \bibfield  {author} {\bibinfo {author} {\bibfnamefont {S.}~\bibnamefont
  {Figge}}, \bibinfo {author} {\bibfnamefont {R.}~\bibnamefont {Kr{\"o}ger}},
  \bibinfo {author} {\bibfnamefont {T.}~\bibnamefont {B{\"o}ttcher}}, \bibinfo
  {author} {\bibfnamefont {P.~L.}\ \bibnamefont {Ryder}},\ and\ \bibinfo
  {author} {\bibfnamefont {D.}~\bibnamefont {Hommel}},\ }\bibfield  {title}
  {\enquote {\bibinfo {title} {Magnesium segregation and the formation of
  pyramidal defects in p-{{GaN}}},}\ }\href {https://doi.org/10.1063/1.1527981}
  {\bibfield  {journal} {\bibinfo  {journal} {Appl. Phys. Lett.}\ }\textbf
  {\bibinfo {volume} {81}},\ \bibinfo {pages} {4748--4750} (\bibinfo {year}
  {2002})}\BibitemShut {NoStop}%
\bibitem [{\citenamefont {Ohba}\ and\ \citenamefont
  {Hatano}(1994)}]{ohbaStudyStrongMemory1994}%
  \BibitemOpen
  \bibfield  {author} {\bibinfo {author} {\bibfnamefont {Y.}~\bibnamefont
  {Ohba}}\ and\ \bibinfo {author} {\bibfnamefont {A.}~\bibnamefont {Hatano}},\
  }\bibfield  {title} {\enquote {\bibinfo {title} {A study on strong memory
  effects for {{Mg}} doping in {{GaN}} metalorganic chemical vapor
  deposition},}\ }\href {https://doi.org/10.1016/0022-0248(94)91053-7}
  {\bibfield  {journal} {\bibinfo  {journal} {J. Cryst. Growth}\ }\textbf
  {\bibinfo {volume} {145}},\ \bibinfo {pages} {214--218} (\bibinfo {year}
  {1994})}\BibitemShut {NoStop}%
\bibitem [{\citenamefont {Kozodoy}, \citenamefont {DenBaars},\ and\
  \citenamefont {Mishra}(2000)}]{kozodoyDepletionRegionEffects2000}%
  \BibitemOpen
  \bibfield  {author} {\bibinfo {author} {\bibfnamefont {P.}~\bibnamefont
  {Kozodoy}}, \bibinfo {author} {\bibfnamefont {S.~P.}\ \bibnamefont
  {DenBaars}},\ and\ \bibinfo {author} {\bibfnamefont {U.~K.}\ \bibnamefont
  {Mishra}},\ }\bibfield  {title} {\enquote {\bibinfo {title} {Depletion region
  effects in {{Mg-doped GaN}}},}\ }\href {https://doi.org/10.1063/1.371939}
  {\bibfield  {journal} {\bibinfo  {journal} {J. Appl. Phys.}\ }\textbf
  {\bibinfo {volume} {87}},\ \bibinfo {pages} {770--775} (\bibinfo {year}
  {2000})}\BibitemShut {NoStop}%
\bibitem [{\citenamefont {Martens}\ \emph {et~al.}(2017)\citenamefont
  {Martens}, \citenamefont {Kuhn}, \citenamefont {Simoneit}, \citenamefont
  {Hagedorn}, \citenamefont {Knauer}, \citenamefont {Wernicke}, \citenamefont
  {Weyers},\ and\ \citenamefont {Kneissl}}]{martensEffectsMagnesiumDoping2017}%
  \BibitemOpen
  \bibfield  {author} {\bibinfo {author} {\bibfnamefont {M.}~\bibnamefont
  {Martens}}, \bibinfo {author} {\bibfnamefont {C.}~\bibnamefont {Kuhn}},
  \bibinfo {author} {\bibfnamefont {T.}~\bibnamefont {Simoneit}}, \bibinfo
  {author} {\bibfnamefont {S.}~\bibnamefont {Hagedorn}}, \bibinfo {author}
  {\bibfnamefont {A.}~\bibnamefont {Knauer}}, \bibinfo {author} {\bibfnamefont
  {T.}~\bibnamefont {Wernicke}}, \bibinfo {author} {\bibfnamefont
  {M.}~\bibnamefont {Weyers}},\ and\ \bibinfo {author} {\bibfnamefont
  {M.}~\bibnamefont {Kneissl}},\ }\bibfield  {title} {\enquote {\bibinfo
  {title} {The effects of magnesium doping on the modal loss in {{AlGaN-based}}
  deep {{UV}} lasers},}\ }\href {https://doi.org/10.1063/1.4977029} {\bibfield
  {journal} {\bibinfo  {journal} {Appl. Phys. Lett.}\ }\textbf {\bibinfo
  {volume} {110}},\ \bibinfo {pages} {081103} (\bibinfo {year}
  {2017})}\BibitemShut {NoStop}%
\bibitem [{\citenamefont {Zhang}\ \emph
  {et~al.}(2022{\natexlab{a}})\citenamefont {Zhang}, \citenamefont {Kushimoto},
  \citenamefont {Yoshikawa}, \citenamefont {Aoto}, \citenamefont {Sasaoka},
  \citenamefont {Schowalter},\ and\ \citenamefont
  {Amano}}]{zhangKeyTemperaturedependentCharacteristics2022}%
  \BibitemOpen
  \bibfield  {author} {\bibinfo {author} {\bibfnamefont {Z.}~\bibnamefont
  {Zhang}}, \bibinfo {author} {\bibfnamefont {M.}~\bibnamefont {Kushimoto}},
  \bibinfo {author} {\bibfnamefont {A.}~\bibnamefont {Yoshikawa}}, \bibinfo
  {author} {\bibfnamefont {K.}~\bibnamefont {Aoto}}, \bibinfo {author}
  {\bibfnamefont {C.}~\bibnamefont {Sasaoka}}, \bibinfo {author} {\bibfnamefont
  {L.~J.}\ \bibnamefont {Schowalter}},\ and\ \bibinfo {author} {\bibfnamefont
  {H.}~\bibnamefont {Amano}},\ }\bibfield  {title} {\enquote {\bibinfo {title}
  {Key temperature-dependent characteristics of {{AlGaN-based UV-C}} laser
  diode and demonstration of room-temperature continuous-wave lasing},}\ }\href
  {https://doi.org/10.1063/5.0124480} {\bibfield  {journal} {\bibinfo
  {journal} {Appl. Phys. Lett.}\ }\textbf {\bibinfo {volume} {121}},\ \bibinfo
  {pages} {222103} (\bibinfo {year} {2022}{\natexlab{a}})}\BibitemShut
  {NoStop}%
\bibitem [{\citenamefont {Zhang}\ \emph
  {et~al.}(2022{\natexlab{b}})\citenamefont {Zhang}, \citenamefont {Kushimoto},
  \citenamefont {Yoshikawa}, \citenamefont {Aoto}, \citenamefont {Schowalter},
  \citenamefont {Sasaoka},\ and\ \citenamefont
  {Amano}}]{zhangContinuouswaveLasingAlGaNbased2022}%
  \BibitemOpen
  \bibfield  {author} {\bibinfo {author} {\bibfnamefont {Z.}~\bibnamefont
  {Zhang}}, \bibinfo {author} {\bibfnamefont {M.}~\bibnamefont {Kushimoto}},
  \bibinfo {author} {\bibfnamefont {A.}~\bibnamefont {Yoshikawa}}, \bibinfo
  {author} {\bibfnamefont {K.}~\bibnamefont {Aoto}}, \bibinfo {author}
  {\bibfnamefont {L.~J.}\ \bibnamefont {Schowalter}}, \bibinfo {author}
  {\bibfnamefont {C.}~\bibnamefont {Sasaoka}},\ and\ \bibinfo {author}
  {\bibfnamefont {H.}~\bibnamefont {Amano}},\ }\bibfield  {title} {\enquote
  {\bibinfo {title} {Continuous-wave lasing of {{AlGaN-based}} ultraviolet
  laser diode at 274.8 nm by current injection},}\ }\href
  {https://doi.org/10.35848/1882-0786/ac6198} {\bibfield  {journal} {\bibinfo
  {journal} {Appl. Phys. Express}\ }\textbf {\bibinfo {volume} {15}},\ \bibinfo
  {pages} {041007} (\bibinfo {year} {2022}{\natexlab{b}})}\BibitemShut
  {NoStop}%
\bibitem [{\citenamefont {Zhang}\ \emph {et~al.}(2019)\citenamefont {Zhang},
  \citenamefont {Kushimoto}, \citenamefont {Sakai}, \citenamefont {Sugiyama},
  \citenamefont {Schowalter}, \citenamefont {Sasaoka},\ and\ \citenamefont
  {Amano}}]{zhang271NmDeepultraviolet2019}%
  \BibitemOpen
  \bibfield  {author} {\bibinfo {author} {\bibfnamefont {Z.}~\bibnamefont
  {Zhang}}, \bibinfo {author} {\bibfnamefont {M.}~\bibnamefont {Kushimoto}},
  \bibinfo {author} {\bibfnamefont {T.}~\bibnamefont {Sakai}}, \bibinfo
  {author} {\bibfnamefont {N.}~\bibnamefont {Sugiyama}}, \bibinfo {author}
  {\bibfnamefont {L.~J.}\ \bibnamefont {Schowalter}}, \bibinfo {author}
  {\bibfnamefont {C.}~\bibnamefont {Sasaoka}},\ and\ \bibinfo {author}
  {\bibfnamefont {H.}~\bibnamefont {Amano}},\ }\bibfield  {title} {\enquote
  {\bibinfo {title} {A 271.8 nm deep-ultraviolet laser diode for room
  temperature operation},}\ }\href {https://doi.org/10.7567/1882-0786/ab50e0}
  {\bibfield  {journal} {\bibinfo  {journal} {Appl. Phys. Express}\ }\textbf
  {\bibinfo {volume} {12}},\ \bibinfo {pages} {124003} (\bibinfo {year}
  {2019})}\BibitemShut {NoStop}%
\bibitem [{\citenamefont {Tanaka}\ \emph {et~al.}(2021)\citenamefont {Tanaka},
  \citenamefont {Ogino}, \citenamefont {Yamada}, \citenamefont {Ogura},
  \citenamefont {Teramura}, \citenamefont {Shimokawa}, \citenamefont
  {Ishizuka}, \citenamefont {Iwayama}, \citenamefont {Sato}, \citenamefont
  {Miyake}, \citenamefont {Iwaya}, \citenamefont {Takeuchi},\ and\
  \citenamefont {Kamiyama}}]{tanakaLowthresholdcurrent85MA2021}%
  \BibitemOpen
  \bibfield  {author} {\bibinfo {author} {\bibfnamefont {S.}~\bibnamefont
  {Tanaka}}, \bibinfo {author} {\bibfnamefont {Y.}~\bibnamefont {Ogino}},
  \bibinfo {author} {\bibfnamefont {K.}~\bibnamefont {Yamada}}, \bibinfo
  {author} {\bibfnamefont {R.}~\bibnamefont {Ogura}}, \bibinfo {author}
  {\bibfnamefont {S.}~\bibnamefont {Teramura}}, \bibinfo {author}
  {\bibfnamefont {M.}~\bibnamefont {Shimokawa}}, \bibinfo {author}
  {\bibfnamefont {S.}~\bibnamefont {Ishizuka}}, \bibinfo {author}
  {\bibfnamefont {S.}~\bibnamefont {Iwayama}}, \bibinfo {author} {\bibfnamefont
  {K.}~\bibnamefont {Sato}}, \bibinfo {author} {\bibfnamefont {H.}~\bibnamefont
  {Miyake}}, \bibinfo {author} {\bibfnamefont {M.}~\bibnamefont {Iwaya}},
  \bibinfo {author} {\bibfnamefont {T.}~\bibnamefont {Takeuchi}},\ and\
  \bibinfo {author} {\bibfnamefont {S.}~\bibnamefont {Kamiyama}},\ }\bibfield
  {title} {\enquote {\bibinfo {title} {Low-threshold-current ( 85 {{mA}}) of
  {{AlGaN-based UV-B}} laser diode with refractive-index waveguide
  structure},}\ }\href {https://doi.org/10.35848/1882-0786/ac200b} {\bibfield
  {journal} {\bibinfo  {journal} {Appl. Phys. Express}\ }\textbf {\bibinfo
  {volume} {14}},\ \bibinfo {pages} {094009} (\bibinfo {year}
  {2021})}\BibitemShut {NoStop}%
\bibitem [{\citenamefont {Lee}\ \emph {et~al.}(2019)\citenamefont {Lee},
  \citenamefont {Bharadwaj}, \citenamefont {Protasenko}, \citenamefont {Xing},\
  and\ \citenamefont {Jena}}]{leeEfficientInGaNPContacts2019}%
  \BibitemOpen
  \bibfield  {author} {\bibinfo {author} {\bibfnamefont {K.}~\bibnamefont
  {Lee}}, \bibinfo {author} {\bibfnamefont {S.}~\bibnamefont {Bharadwaj}},
  \bibinfo {author} {\bibfnamefont {V.}~\bibnamefont {Protasenko}}, \bibinfo
  {author} {\bibfnamefont {H.}~\bibnamefont {Xing}},\ and\ \bibinfo {author}
  {\bibfnamefont {D.}~\bibnamefont {Jena}},\ }\bibfield  {title} {\enquote
  {\bibinfo {title} {Efficient {{InGaN}} p-{{Contacts}} for deep-{{UV Light
  Emitting Diodes}}},}\ }in\ \href
  {https://doi.org/10.1109/DRC46940.2019.9046469} {\emph {\bibinfo {booktitle}
  {2019 {{Device Research Conference}} ({{DRC}})}}}\ (\bibinfo {year} {2019})\
  pp.\ \bibinfo {pages} {171--172}\BibitemShut {NoStop}%
\bibitem [{\citenamefont {Kolbe}\ \emph {et~al.}(2023)\citenamefont {Kolbe},
  \citenamefont {Knauer}, \citenamefont {Rass}, \citenamefont {Cho},
  \citenamefont {Hagedorn}, \citenamefont {Bilchenko}, \citenamefont {Muhin},
  \citenamefont {Ruschel}, \citenamefont {Kneissl}, \citenamefont {Einfeldt},\
  and\ \citenamefont {Weyers}}]{kolbe234NmFarultravioletC2023}%
  \BibitemOpen
  \bibfield  {author} {\bibinfo {author} {\bibfnamefont {T.}~\bibnamefont
  {Kolbe}}, \bibinfo {author} {\bibfnamefont {A.}~\bibnamefont {Knauer}},
  \bibinfo {author} {\bibfnamefont {J.}~\bibnamefont {Rass}}, \bibinfo {author}
  {\bibfnamefont {H.~K.}\ \bibnamefont {Cho}}, \bibinfo {author} {\bibfnamefont
  {S.}~\bibnamefont {Hagedorn}}, \bibinfo {author} {\bibfnamefont
  {F.}~\bibnamefont {Bilchenko}}, \bibinfo {author} {\bibfnamefont
  {A.}~\bibnamefont {Muhin}}, \bibinfo {author} {\bibfnamefont
  {J.}~\bibnamefont {Ruschel}}, \bibinfo {author} {\bibfnamefont
  {M.}~\bibnamefont {Kneissl}}, \bibinfo {author} {\bibfnamefont
  {S.}~\bibnamefont {Einfeldt}},\ and\ \bibinfo {author} {\bibfnamefont
  {M.}~\bibnamefont {Weyers}},\ }\bibfield  {title} {\enquote {\bibinfo {title}
  {234\,nm far-ultraviolet-{{C}} light-emitting diodes with polarization-doped
  hole injection layer},}\ }\href {https://doi.org/10.1063/5.0143661}
  {\bibfield  {journal} {\bibinfo  {journal} {Appl. Phys. Lett.}\ }\textbf
  {\bibinfo {volume} {122}},\ \bibinfo {pages} {191101} (\bibinfo {year}
  {2023})}\BibitemShut {NoStop}%
\bibitem [{\citenamefont {Kumabe}\ \emph {et~al.}(2022)\citenamefont {Kumabe},
  \citenamefont {Kawasaki}, \citenamefont {Watanabe}, \citenamefont {Nitta},
  \citenamefont {Honda},\ and\ \citenamefont
  {Amano}}]{kumabeSpaceChargeProfiles2022}%
  \BibitemOpen
  \bibfield  {author} {\bibinfo {author} {\bibfnamefont {T.}~\bibnamefont
  {Kumabe}}, \bibinfo {author} {\bibfnamefont {S.}~\bibnamefont {Kawasaki}},
  \bibinfo {author} {\bibfnamefont {H.}~\bibnamefont {Watanabe}}, \bibinfo
  {author} {\bibfnamefont {S.}~\bibnamefont {Nitta}}, \bibinfo {author}
  {\bibfnamefont {Y.}~\bibnamefont {Honda}},\ and\ \bibinfo {author}
  {\bibfnamefont {H.}~\bibnamefont {Amano}},\ }\bibfield  {title} {\enquote
  {\bibinfo {title} {Space\textendash{{Charge Profiles}} and {{Carrier
  Transport Properties}} in {{Dopant-Free GaN-Based}} p-n {{Junction Formed}}
  by {{Distributed Polarization Doping}}},}\ }\href
  {https://doi.org/10.1002/pssr.202200127} {\bibfield  {journal} {\bibinfo
  {journal} {Phys. Status Solidi Rapid Res. Lett.}\ }\textbf {\bibinfo {volume}
  {16}},\ \bibinfo {pages} {2200127} (\bibinfo {year} {2022})}\BibitemShut
  {NoStop}%
\bibitem [{\citenamefont {Nomoto}\ \emph {et~al.}(2022)\citenamefont {Nomoto},
  \citenamefont {Li}, \citenamefont {Song}, \citenamefont {Hu}, \citenamefont
  {Zhu}, \citenamefont {Qi}, \citenamefont {Protasenko}, \citenamefont {Zhang},
  \citenamefont {Pan}, \citenamefont {Gao}, \citenamefont {Marchand},
  \citenamefont {Johnson}, \citenamefont {Jena},\ and\ \citenamefont
  {Xing}}]{nomotoDistributedPolarizationdopedGaN2022}%
  \BibitemOpen
  \bibfield  {author} {\bibinfo {author} {\bibfnamefont {K.}~\bibnamefont
  {Nomoto}}, \bibinfo {author} {\bibfnamefont {W.}~\bibnamefont {Li}}, \bibinfo
  {author} {\bibfnamefont {B.}~\bibnamefont {Song}}, \bibinfo {author}
  {\bibfnamefont {Z.}~\bibnamefont {Hu}}, \bibinfo {author} {\bibfnamefont
  {M.}~\bibnamefont {Zhu}}, \bibinfo {author} {\bibfnamefont {M.}~\bibnamefont
  {Qi}}, \bibinfo {author} {\bibfnamefont {V.}~\bibnamefont {Protasenko}},
  \bibinfo {author} {\bibfnamefont {Z.}~\bibnamefont {Zhang}}, \bibinfo
  {author} {\bibfnamefont {M.}~\bibnamefont {Pan}}, \bibinfo {author}
  {\bibfnamefont {X.}~\bibnamefont {Gao}}, \bibinfo {author} {\bibfnamefont
  {H.}~\bibnamefont {Marchand}}, \bibinfo {author} {\bibfnamefont
  {W.}~\bibnamefont {Johnson}}, \bibinfo {author} {\bibfnamefont
  {D.}~\bibnamefont {Jena}},\ and\ \bibinfo {author} {\bibfnamefont {H.~G.}\
  \bibnamefont {Xing}},\ }\bibfield  {title} {\enquote {\bibinfo {title}
  {Distributed polarization-doped {{GaN}} p\textendash n diodes with near-unity
  ideality factor and avalanche breakdown voltage of 1.25 {{kV}}},}\ }\href
  {https://doi.org/10.1063/5.0083302} {\bibfield  {journal} {\bibinfo
  {journal} {Appl. Phys. Lett.}\ }\textbf {\bibinfo {volume} {120}},\ \bibinfo
  {pages} {122111} (\bibinfo {year} {2022})}\BibitemShut {NoStop}%
\bibitem [{\citenamefont {Hu}\ \emph {et~al.}(2015)\citenamefont {Hu},
  \citenamefont {Nomoto}, \citenamefont {Song}, \citenamefont {Zhu},
  \citenamefont {Qi}, \citenamefont {Pan}, \citenamefont {Gao}, \citenamefont
  {Protasenko}, \citenamefont {Jena},\ and\ \citenamefont
  {Xing}}]{huUnityIdealityFactor2015}%
  \BibitemOpen
  \bibfield  {author} {\bibinfo {author} {\bibfnamefont {Z.}~\bibnamefont
  {Hu}}, \bibinfo {author} {\bibfnamefont {K.}~\bibnamefont {Nomoto}}, \bibinfo
  {author} {\bibfnamefont {B.}~\bibnamefont {Song}}, \bibinfo {author}
  {\bibfnamefont {M.}~\bibnamefont {Zhu}}, \bibinfo {author} {\bibfnamefont
  {M.}~\bibnamefont {Qi}}, \bibinfo {author} {\bibfnamefont {M.}~\bibnamefont
  {Pan}}, \bibinfo {author} {\bibfnamefont {X.}~\bibnamefont {Gao}}, \bibinfo
  {author} {\bibfnamefont {V.}~\bibnamefont {Protasenko}}, \bibinfo {author}
  {\bibfnamefont {D.}~\bibnamefont {Jena}},\ and\ \bibinfo {author}
  {\bibfnamefont {H.~G.}\ \bibnamefont {Xing}},\ }\bibfield  {title} {\enquote
  {\bibinfo {title} {Near unity ideality factor and {{Shockley-Read-Hall}}
  lifetime in {{GaN-on-GaN}} p-n diodes with avalanche breakdown},}\ }\href
  {https://doi.org/10.1063/1.4937436} {\bibfield  {journal} {\bibinfo
  {journal} {Appl. Phys. Lett.}\ }\textbf {\bibinfo {volume} {107}},\ \bibinfo
  {pages} {243501} (\bibinfo {year} {2015})}\BibitemShut {NoStop}%
\bibitem [{\citenamefont {Li}\ \emph {et~al.}(2012)\citenamefont {Li},
  \citenamefont {Ware}, \citenamefont {Wu}, \citenamefont {Minor},
  \citenamefont {Wang}, \citenamefont {Wu}, \citenamefont {Jiang},\ and\
  \citenamefont {Salamo}}]{liPolarizationInducedPnjunction2012}%
  \BibitemOpen
  \bibfield  {author} {\bibinfo {author} {\bibfnamefont {S.}~\bibnamefont
  {Li}}, \bibinfo {author} {\bibfnamefont {M.}~\bibnamefont {Ware}}, \bibinfo
  {author} {\bibfnamefont {J.}~\bibnamefont {Wu}}, \bibinfo {author}
  {\bibfnamefont {P.}~\bibnamefont {Minor}}, \bibinfo {author} {\bibfnamefont
  {Z.}~\bibnamefont {Wang}}, \bibinfo {author} {\bibfnamefont {Z.}~\bibnamefont
  {Wu}}, \bibinfo {author} {\bibfnamefont {Y.}~\bibnamefont {Jiang}},\ and\
  \bibinfo {author} {\bibfnamefont {G.~J.}\ \bibnamefont {Salamo}},\ }\bibfield
   {title} {\enquote {\bibinfo {title} {Polarization induced pn-junction
  without dopant in graded {{AlGaN}} coherently strained on {{GaN}}},}\ }\href
  {https://doi.org/10.1063/1.4753993} {\bibfield  {journal} {\bibinfo
  {journal} {Appl. Phys. Lett.}\ }\textbf {\bibinfo {volume} {101}},\ \bibinfo
  {pages} {122103} (\bibinfo {year} {2012})}\BibitemShut {NoStop}%
\bibitem [{\citenamefont {Simon}\ \emph {et~al.}(2010)\citenamefont {Simon},
  \citenamefont {Protasenko}, \citenamefont {Lian}, \citenamefont {Xing},\ and\
  \citenamefont {Jena}}]{simonPolarizationInducedHoleDoping2010}%
  \BibitemOpen
  \bibfield  {author} {\bibinfo {author} {\bibfnamefont {J.}~\bibnamefont
  {Simon}}, \bibinfo {author} {\bibfnamefont {V.}~\bibnamefont {Protasenko}},
  \bibinfo {author} {\bibfnamefont {C.}~\bibnamefont {Lian}}, \bibinfo {author}
  {\bibfnamefont {H.}~\bibnamefont {Xing}},\ and\ \bibinfo {author}
  {\bibfnamefont {D.}~\bibnamefont {Jena}},\ }\bibfield  {title} {\enquote
  {\bibinfo {title} {Polarization-{{Induced Hole Doping}} in
  {{Wide}}\textendash{{Band-Gap Uniaxial Semiconductor Heterostructures}}},}\
  }\href {https://doi.org/10.1126/science.1183226} {\bibfield  {journal}
  {\bibinfo  {journal} {Science}\ }\textbf {\bibinfo {volume} {327}},\ \bibinfo
  {pages} {60--64} (\bibinfo {year} {2010})}\BibitemShut {NoStop}%
\bibitem [{\citenamefont {Ahmad}\ \emph {et~al.}(2022)\citenamefont {Ahmad},
  \citenamefont {Strak}, \citenamefont {Kempisty}, \citenamefont {Sakowski},
  \citenamefont {Piechota}, \citenamefont {Kangawa}, \citenamefont {Grzegory},
  \citenamefont {Leszczynski}, \citenamefont {Zytkiewicz}, \citenamefont
  {Muziol}, \citenamefont {Monroy}, \citenamefont {Kaminska},\ and\
  \citenamefont {Krukowski}}]{ahmadPolarizationDopingInitio2022}%
  \BibitemOpen
  \bibfield  {author} {\bibinfo {author} {\bibfnamefont {A.}~\bibnamefont
  {Ahmad}}, \bibinfo {author} {\bibfnamefont {P.}~\bibnamefont {Strak}},
  \bibinfo {author} {\bibfnamefont {P.}~\bibnamefont {Kempisty}}, \bibinfo
  {author} {\bibfnamefont {K.}~\bibnamefont {Sakowski}}, \bibinfo {author}
  {\bibfnamefont {J.}~\bibnamefont {Piechota}}, \bibinfo {author}
  {\bibfnamefont {Y.}~\bibnamefont {Kangawa}}, \bibinfo {author} {\bibfnamefont
  {I.}~\bibnamefont {Grzegory}}, \bibinfo {author} {\bibfnamefont
  {M.}~\bibnamefont {Leszczynski}}, \bibinfo {author} {\bibfnamefont {Z.~R.}\
  \bibnamefont {Zytkiewicz}}, \bibinfo {author} {\bibfnamefont
  {G.}~\bibnamefont {Muziol}}, \bibinfo {author} {\bibfnamefont
  {E.}~\bibnamefont {Monroy}}, \bibinfo {author} {\bibfnamefont
  {A.}~\bibnamefont {Kaminska}},\ and\ \bibinfo {author} {\bibfnamefont
  {S.}~\bibnamefont {Krukowski}},\ }\bibfield  {title} {\enquote {\bibinfo
  {title} {Polarization doping\textemdash{{Ab}} initio verification of the
  concept: {{Charge}} conservation and nonlocality},}\ }\href
  {https://doi.org/10.1063/5.0098909} {\bibfield  {journal} {\bibinfo
  {journal} {J. Appl. Phys.}\ }\textbf {\bibinfo {volume} {132}},\ \bibinfo
  {pages} {064301} (\bibinfo {year} {2022})}\BibitemShut {NoStop}%
\bibitem [{\citenamefont {Chaudhuri}\ \emph {et~al.}(2019)\citenamefont
  {Chaudhuri}, \citenamefont {Bader}, \citenamefont {Chen}, \citenamefont
  {Muller}, \citenamefont {Xing},\ and\ \citenamefont
  {Jena}}]{chaudhuriPolarizationinduced2DHole2019}%
  \BibitemOpen
  \bibfield  {author} {\bibinfo {author} {\bibfnamefont {R.}~\bibnamefont
  {Chaudhuri}}, \bibinfo {author} {\bibfnamefont {S.~J.}\ \bibnamefont
  {Bader}}, \bibinfo {author} {\bibfnamefont {Z.}~\bibnamefont {Chen}},
  \bibinfo {author} {\bibfnamefont {D.~A.}\ \bibnamefont {Muller}}, \bibinfo
  {author} {\bibfnamefont {H.~G.}\ \bibnamefont {Xing}},\ and\ \bibinfo
  {author} {\bibfnamefont {D.}~\bibnamefont {Jena}},\ }\bibfield  {title}
  {\enquote {\bibinfo {title} {A polarization-induced {{2D}} hole gas in
  undoped gallium nitride quantum wells},}\ }\href
  {https://doi.org/10.1126/science.aau8623} {\bibfield  {journal} {\bibinfo
  {journal} {Science}\ }\textbf {\bibinfo {volume} {365}},\ \bibinfo {pages}
  {1454--1457} (\bibinfo {year} {2019})}\BibitemShut {NoStop}%
\bibitem [{\citenamefont {Zhang}\ \emph {et~al.}(2020)\citenamefont {Zhang},
  \citenamefont {Kushimoto}, \citenamefont {Horita}, \citenamefont {Sugiyama},
  \citenamefont {Schowalter}, \citenamefont {Sasaoka},\ and\ \citenamefont
  {Amano}}]{zhangSpaceChargeProfile2020}%
  \BibitemOpen
  \bibfield  {author} {\bibinfo {author} {\bibfnamefont {Z.}~\bibnamefont
  {Zhang}}, \bibinfo {author} {\bibfnamefont {M.}~\bibnamefont {Kushimoto}},
  \bibinfo {author} {\bibfnamefont {M.}~\bibnamefont {Horita}}, \bibinfo
  {author} {\bibfnamefont {N.}~\bibnamefont {Sugiyama}}, \bibinfo {author}
  {\bibfnamefont {L.~J.}\ \bibnamefont {Schowalter}}, \bibinfo {author}
  {\bibfnamefont {C.}~\bibnamefont {Sasaoka}},\ and\ \bibinfo {author}
  {\bibfnamefont {H.}~\bibnamefont {Amano}},\ }\bibfield  {title} {\enquote
  {\bibinfo {title} {Space charge profile study of {{AlGaN-based}} p-type
  distributed polarization doped claddings without impurity doping for {{UV-C}}
  laser diodes},}\ }\href {https://doi.org/10.1063/5.0027789} {\bibfield
  {journal} {\bibinfo  {journal} {Appl. Phys. Lett.}\ }\textbf {\bibinfo
  {volume} {117}},\ \bibinfo {pages} {152104} (\bibinfo {year}
  {2020})}\BibitemShut {NoStop}%
\bibitem [{\citenamefont {Zhang}\ \emph {et~al.}(2021)\citenamefont {Zhang},
  \citenamefont {Encomendero}, \citenamefont {Chaudhuri}, \citenamefont {Cho},
  \citenamefont {Protasenko}, \citenamefont {Nomoto}, \citenamefont {Lee},
  \citenamefont {Toita}, \citenamefont {Xing},\ and\ \citenamefont
  {Jena}}]{zhangPolarizationinduced2DHole2021}%
  \BibitemOpen
  \bibfield  {author} {\bibinfo {author} {\bibfnamefont {Z.}~\bibnamefont
  {Zhang}}, \bibinfo {author} {\bibfnamefont {J.}~\bibnamefont {Encomendero}},
  \bibinfo {author} {\bibfnamefont {R.}~\bibnamefont {Chaudhuri}}, \bibinfo
  {author} {\bibfnamefont {Y.}~\bibnamefont {Cho}}, \bibinfo {author}
  {\bibfnamefont {V.}~\bibnamefont {Protasenko}}, \bibinfo {author}
  {\bibfnamefont {K.}~\bibnamefont {Nomoto}}, \bibinfo {author} {\bibfnamefont
  {K.}~\bibnamefont {Lee}}, \bibinfo {author} {\bibfnamefont {M.}~\bibnamefont
  {Toita}}, \bibinfo {author} {\bibfnamefont {H.~G.}\ \bibnamefont {Xing}},\
  and\ \bibinfo {author} {\bibfnamefont {D.}~\bibnamefont {Jena}},\ }\bibfield
  {title} {\enquote {\bibinfo {title} {Polarization-induced {{2D}} hole gases
  in pseudomorphic undoped {{GaN}}/{{AlN}} heterostructures on single-crystal
  {{AlN}} substrates},}\ }\href {https://doi.org/10.1063/5.0066072} {\bibfield
  {journal} {\bibinfo  {journal} {Appl. Phys. Lett.}\ }\textbf {\bibinfo
  {volume} {119}},\ \bibinfo {pages} {162104} (\bibinfo {year}
  {2021})}\BibitemShut {NoStop}%
\bibitem [{\citenamefont {Cho}\ \emph {et~al.}(2020)\citenamefont {Cho},
  \citenamefont {Chang}, \citenamefont {Lee}, \citenamefont {Gong},
  \citenamefont {Nomoto}, \citenamefont {Toita}, \citenamefont {Schowalter},
  \citenamefont {Muller}, \citenamefont {Jena},\ and\ \citenamefont
  {Xing}}]{choMolecularBeamHomoepitaxy2020}%
  \BibitemOpen
  \bibfield  {author} {\bibinfo {author} {\bibfnamefont {Y.}~\bibnamefont
  {Cho}}, \bibinfo {author} {\bibfnamefont {C.~S.}\ \bibnamefont {Chang}},
  \bibinfo {author} {\bibfnamefont {K.}~\bibnamefont {Lee}}, \bibinfo {author}
  {\bibfnamefont {M.}~\bibnamefont {Gong}}, \bibinfo {author} {\bibfnamefont
  {K.}~\bibnamefont {Nomoto}}, \bibinfo {author} {\bibfnamefont
  {M.}~\bibnamefont {Toita}}, \bibinfo {author} {\bibfnamefont {L.~J.}\
  \bibnamefont {Schowalter}}, \bibinfo {author} {\bibfnamefont {D.~A.}\
  \bibnamefont {Muller}}, \bibinfo {author} {\bibfnamefont {D.}~\bibnamefont
  {Jena}},\ and\ \bibinfo {author} {\bibfnamefont {H.~G.}\ \bibnamefont
  {Xing}},\ }\bibfield  {title} {\enquote {\bibinfo {title} {Molecular beam
  homoepitaxy on bulk {{AlN}} enabled by aluminum-assisted surface cleaning},}\
  }\href {https://doi.org/10.1063/1.5143968} {\bibfield  {journal} {\bibinfo
  {journal} {Appl. Phys. Lett.}\ }\textbf {\bibinfo {volume} {116}},\ \bibinfo
  {pages} {172106} (\bibinfo {year} {2020})}\BibitemShut {NoStop}%
\bibitem [{\citenamefont {Lee}\ \emph {et~al.}(2020)\citenamefont {Lee},
  \citenamefont {Cho}, \citenamefont {Schowalter}, \citenamefont {Toita},
  \citenamefont {Xing},\ and\ \citenamefont {Jena}}]{leeSurfaceControlMBE2020}%
  \BibitemOpen
  \bibfield  {author} {\bibinfo {author} {\bibfnamefont {K.}~\bibnamefont
  {Lee}}, \bibinfo {author} {\bibfnamefont {Y.}~\bibnamefont {Cho}}, \bibinfo
  {author} {\bibfnamefont {L.~J.}\ \bibnamefont {Schowalter}}, \bibinfo
  {author} {\bibfnamefont {M.}~\bibnamefont {Toita}}, \bibinfo {author}
  {\bibfnamefont {H.~G.}\ \bibnamefont {Xing}},\ and\ \bibinfo {author}
  {\bibfnamefont {D.}~\bibnamefont {Jena}},\ }\bibfield  {title} {\enquote
  {\bibinfo {title} {Surface control and {{MBE}} growth diagram for homoepitaxy
  on single-crystal {{AlN}} substrates},}\ }\href
  {https://doi.org/10.1063/5.0010813} {\bibfield  {journal} {\bibinfo
  {journal} {Appl. Phys. Lett.}\ }\textbf {\bibinfo {volume} {116}},\ \bibinfo
  {pages} {262102} (\bibinfo {year} {2020})}\BibitemShut {NoStop}%
\bibitem [{\citenamefont {Sze}, \citenamefont {Li},\ and\ \citenamefont
  {Ng}(2021)}]{szePhysicsSemiconductorDevices2021}%
  \BibitemOpen
  \bibfield  {author} {\bibinfo {author} {\bibfnamefont {S.~M.}\ \bibnamefont
  {Sze}}, \bibinfo {author} {\bibfnamefont {Y.}~\bibnamefont {Li}},\ and\
  \bibinfo {author} {\bibfnamefont {K.~K.}\ \bibnamefont {Ng}},\ }\href@noop {}
  {\emph {\bibinfo {title} {Physics of {{Semiconductor Devices}}}}}\ (\bibinfo
  {publisher} {{John Wiley \& Sons}},\ \bibinfo {year} {2021})\BibitemShut
  {NoStop}%
\bibitem [{\citenamefont {Shah}\ \emph {et~al.}(2003)\citenamefont {Shah},
  \citenamefont {Li}, \citenamefont {Gessmann},\ and\ \citenamefont
  {Schubert}}]{shahExperimentalAnalysisTheoretical2003}%
  \BibitemOpen
  \bibfield  {author} {\bibinfo {author} {\bibfnamefont {J.~M.}\ \bibnamefont
  {Shah}}, \bibinfo {author} {\bibfnamefont {Y.-L.}\ \bibnamefont {Li}},
  \bibinfo {author} {\bibfnamefont {{\relax Th}.}~\bibnamefont {Gessmann}},\
  and\ \bibinfo {author} {\bibfnamefont {E.~F.}\ \bibnamefont {Schubert}},\
  }\bibfield  {title} {\enquote {\bibinfo {title} {Experimental analysis and
  theoretical model for anomalously high ideality factors (n{$\gg$}2.0) in
  {{AlGaN}}/{{GaN}} p-n junction diodes},}\ }\href
  {https://doi.org/10.1063/1.1593218} {\bibfield  {journal} {\bibinfo
  {journal} {J. Appl. Phys.}\ }\textbf {\bibinfo {volume} {94}},\ \bibinfo
  {pages} {2627--2630} (\bibinfo {year} {2003})}\BibitemShut {NoStop}%
\bibitem [{\citenamefont {Ferdous}\ \emph {et~al.}(2007)\citenamefont
  {Ferdous}, \citenamefont {Wang}, \citenamefont {Fairchild},\ and\
  \citenamefont {Hersee}}]{ferdousEffectThreadingDefects2007}%
  \BibitemOpen
  \bibfield  {author} {\bibinfo {author} {\bibfnamefont {M.~S.}\ \bibnamefont
  {Ferdous}}, \bibinfo {author} {\bibfnamefont {X.}~\bibnamefont {Wang}},
  \bibinfo {author} {\bibfnamefont {M.~N.}\ \bibnamefont {Fairchild}},\ and\
  \bibinfo {author} {\bibfnamefont {S.~D.}\ \bibnamefont {Hersee}},\ }\bibfield
   {title} {\enquote {\bibinfo {title} {Effect of threading defects on
  {{InGaN}}/{{GaN}} multiple quantum well light emitting diodes},}\ }\href
  {https://doi.org/10.1063/1.2822395} {\bibfield  {journal} {\bibinfo
  {journal} {Appl. Phys. Lett.}\ }\textbf {\bibinfo {volume} {91}},\ \bibinfo
  {pages} {231107} (\bibinfo {year} {2007})}\BibitemShut {NoStop}%
\bibitem [{\citenamefont
  {Mott}(1969)}]{mottConductionNoncrystallineMaterials1969}%
  \BibitemOpen
  \bibfield  {author} {\bibinfo {author} {\bibfnamefont {N.~F.}\ \bibnamefont
  {Mott}},\ }\bibfield  {title} {\enquote {\bibinfo {title} {Conduction in
  non-crystalline materials},}\ }\href
  {https://doi.org/10.1080/14786436908216338} {\bibfield  {journal} {\bibinfo
  {journal} {The Philosophical Magazine: A Journal of Theoretical Experimental
  and Applied Physics}\ }\textbf {\bibinfo {volume} {19}},\ \bibinfo {pages}
  {835--852} (\bibinfo {year} {1969})}\BibitemShut {NoStop}%
\bibitem [{\citenamefont
  {Stauffer}(2008)}]{staufferFundamentalsSemiconductorCV2008}%
  \BibitemOpen
  \bibfield  {author} {\bibinfo {author} {\bibfnamefont {L.}~\bibnamefont
  {Stauffer}},\ }\bibfield  {title} {\enquote {\bibinfo {title} {Fundamentals
  of semiconductor {{C-V}} measurements.}}\ }\href@noop {} {\bibfield
  {journal} {\bibinfo  {journal} {EE: Eval. Eng.}\ }\textbf {\bibinfo {volume}
  {47}},\ \bibinfo {pages} {20--24} (\bibinfo {year} {2008})}\BibitemShut
  {NoStop}%
\bibitem [{\citenamefont {Wood}\ and\ \citenamefont
  {Jena}(2007)}]{woodPolarizationEffectsSemiconductors2007}%
  \BibitemOpen
  \bibfield  {author} {\bibinfo {author} {\bibfnamefont {C.}~\bibnamefont
  {Wood}}\ and\ \bibinfo {author} {\bibfnamefont {D.}~\bibnamefont {Jena}},\
  }\href@noop {} {\emph {\bibinfo {title} {Polarization {{Effects}} in
  {{Semiconductors}}: {{From Ab Initio Theory}} to {{Device Applications}}}}}\
  (\bibinfo  {publisher} {{Springer Science \& Business Media}},\ \bibinfo
  {year} {2007})\BibitemShut {NoStop}%
\bibitem [{\citenamefont {Feneberg}\ \emph {et~al.}(2013)\citenamefont
  {Feneberg}, \citenamefont {Romero}, \citenamefont {R{\"o}ppischer},
  \citenamefont {Cobet}, \citenamefont {Esser}, \citenamefont {Neuschl},
  \citenamefont {Thonke}, \citenamefont {Bickermann},\ and\ \citenamefont
  {Goldhahn}}]{fenebergAnisotropicAbsorptionEmission2013}%
  \BibitemOpen
  \bibfield  {author} {\bibinfo {author} {\bibfnamefont {M.}~\bibnamefont
  {Feneberg}}, \bibinfo {author} {\bibfnamefont {M.~F.}\ \bibnamefont
  {Romero}}, \bibinfo {author} {\bibfnamefont {M.}~\bibnamefont
  {R{\"o}ppischer}}, \bibinfo {author} {\bibfnamefont {C.}~\bibnamefont
  {Cobet}}, \bibinfo {author} {\bibfnamefont {N.}~\bibnamefont {Esser}},
  \bibinfo {author} {\bibfnamefont {B.}~\bibnamefont {Neuschl}}, \bibinfo
  {author} {\bibfnamefont {K.}~\bibnamefont {Thonke}}, \bibinfo {author}
  {\bibfnamefont {M.}~\bibnamefont {Bickermann}},\ and\ \bibinfo {author}
  {\bibfnamefont {R.}~\bibnamefont {Goldhahn}},\ }\bibfield  {title} {\enquote
  {\bibinfo {title} {Anisotropic absorption and emission of bulk
  $(1\overline{1}00)$ {{AlN}}},}\ }\href
  {https://doi.org/10.1103/PhysRevB.87.235209} {\bibfield  {journal} {\bibinfo
  {journal} {Phys. Rev. B}\ }\textbf {\bibinfo {volume} {87}},\ \bibinfo
  {pages} {235209} (\bibinfo {year} {2013})}\BibitemShut {NoStop}%
\bibitem [{\citenamefont {Feneberg}\ \emph {et~al.}(2014)\citenamefont
  {Feneberg}, \citenamefont {Osterburg}, \citenamefont {Lange}, \citenamefont
  {Lidig}, \citenamefont {Garke}, \citenamefont {Goldhahn}, \citenamefont
  {Richter}, \citenamefont {Netzel}, \citenamefont {Neumann}, \citenamefont
  {Esser}, \citenamefont {Fritze}, \citenamefont {Witte}, \citenamefont
  {Bl{\"a}sing}, \citenamefont {Dadgar},\ and\ \citenamefont
  {Krost}}]{fenebergBandGapRenormalization2014}%
  \BibitemOpen
  \bibfield  {author} {\bibinfo {author} {\bibfnamefont {M.}~\bibnamefont
  {Feneberg}}, \bibinfo {author} {\bibfnamefont {S.}~\bibnamefont {Osterburg}},
  \bibinfo {author} {\bibfnamefont {K.}~\bibnamefont {Lange}}, \bibinfo
  {author} {\bibfnamefont {C.}~\bibnamefont {Lidig}}, \bibinfo {author}
  {\bibfnamefont {B.}~\bibnamefont {Garke}}, \bibinfo {author} {\bibfnamefont
  {R.}~\bibnamefont {Goldhahn}}, \bibinfo {author} {\bibfnamefont
  {E.}~\bibnamefont {Richter}}, \bibinfo {author} {\bibfnamefont
  {C.}~\bibnamefont {Netzel}}, \bibinfo {author} {\bibfnamefont {M.~D.}\
  \bibnamefont {Neumann}}, \bibinfo {author} {\bibfnamefont {N.}~\bibnamefont
  {Esser}}, \bibinfo {author} {\bibfnamefont {S.}~\bibnamefont {Fritze}},
  \bibinfo {author} {\bibfnamefont {H.}~\bibnamefont {Witte}}, \bibinfo
  {author} {\bibfnamefont {J.}~\bibnamefont {Bl{\"a}sing}}, \bibinfo {author}
  {\bibfnamefont {A.}~\bibnamefont {Dadgar}},\ and\ \bibinfo {author}
  {\bibfnamefont {A.}~\bibnamefont {Krost}},\ }\bibfield  {title} {\enquote
  {\bibinfo {title} {Band gap renormalization and {{Burstein-Moss}} effect in
  silicon- and germanium-doped wurtzite {{GaN}} up to 10$^{20}$ cm$^{-3}$},}\
  }\href {https://doi.org/10.1103/PhysRevB.90.075203} {\bibfield  {journal}
  {\bibinfo  {journal} {Phys. Rev. B}\ }\textbf {\bibinfo {volume} {90}},\
  \bibinfo {pages} {075203} (\bibinfo {year} {2014})}\BibitemShut {NoStop}%
\bibitem [{\citenamefont {Kanisawa}\ \emph {et~al.}(2001)\citenamefont
  {Kanisawa}, \citenamefont {Butcher}, \citenamefont {Yamaguchi},\ and\
  \citenamefont {Hirayama}}]{kanisawaImagingFriedelOscillation2001}%
  \BibitemOpen
  \bibfield  {author} {\bibinfo {author} {\bibfnamefont {K.}~\bibnamefont
  {Kanisawa}}, \bibinfo {author} {\bibfnamefont {M.~J.}\ \bibnamefont
  {Butcher}}, \bibinfo {author} {\bibfnamefont {H.}~\bibnamefont {Yamaguchi}},\
  and\ \bibinfo {author} {\bibfnamefont {Y.}~\bibnamefont {Hirayama}},\
  }\bibfield  {title} {\enquote {\bibinfo {title} {Imaging of {{Friedel
  Oscillation Patterns}} of {{Two-Dimensionally Accumulated Electrons}} at
  {{Epitaxially Grown InAs}}(111) $\mathit{A}$ {{Surfaces}}},}\ }\href
  {https://doi.org/10.1103/PhysRevLett.86.3384} {\bibfield  {journal} {\bibinfo
   {journal} {Phys. Rev. Lett.}\ }\textbf {\bibinfo {volume} {86}},\ \bibinfo
  {pages} {3384--3387} (\bibinfo {year} {2001})}\BibitemShut {NoStop}%
\bibitem [{\citenamefont {{van der Wielen}}, \citenamefont {{van Roij}},\ and\
  \citenamefont {{van
  Kempen}}(1996)}]{vanderwielenDirectObservationFriedel1996}%
  \BibitemOpen
  \bibfield  {author} {\bibinfo {author} {\bibfnamefont {M.~C. M.~M.}\
  \bibnamefont {{van der Wielen}}}, \bibinfo {author} {\bibfnamefont
  {A.~J.~A.}\ \bibnamefont {{van Roij}}},\ and\ \bibinfo {author}
  {\bibfnamefont {H.}~\bibnamefont {{van Kempen}}},\ }\bibfield  {title}
  {\enquote {\bibinfo {title} {Direct {{Observation}} of {{Friedel
  Oscillations}} around {{Incorporated}} {Si}$_{{{\mathrm{Ga}}}}$ {{Dopants}}
  in {{GaAs}} by {{Low-Temperature Scanning Tunneling Microscopy}}},}\ }\href
  {https://doi.org/10.1103/PhysRevLett.76.1075} {\bibfield  {journal} {\bibinfo
   {journal} {Phys. Rev. Lett.}\ }\textbf {\bibinfo {volume} {76}},\ \bibinfo
  {pages} {1075--1078} (\bibinfo {year} {1996})}\BibitemShut {NoStop}%
\bibitem [{SiL()}]{SiLENSe}%
  \BibitemOpen
  \href@noop {} {\enquote {\bibinfo {title} {{{SiLENSe}}},}\ }\BibitemShut
  {NoStop}%
\bibitem [{\citenamefont {Chichibu}\ \emph {et~al.}(2013)\citenamefont
  {Chichibu}, \citenamefont {Miyake}, \citenamefont {Ishikawa}, \citenamefont
  {Tashiro}, \citenamefont {Ohtomo}, \citenamefont {Furusawa}, \citenamefont
  {Hazu}, \citenamefont {Hiramatsu},\ and\ \citenamefont
  {Uedono}}]{chichibuImpactsSidopingResultant2013}%
  \BibitemOpen
  \bibfield  {author} {\bibinfo {author} {\bibfnamefont {S.~F.}\ \bibnamefont
  {Chichibu}}, \bibinfo {author} {\bibfnamefont {H.}~\bibnamefont {Miyake}},
  \bibinfo {author} {\bibfnamefont {Y.}~\bibnamefont {Ishikawa}}, \bibinfo
  {author} {\bibfnamefont {M.}~\bibnamefont {Tashiro}}, \bibinfo {author}
  {\bibfnamefont {T.}~\bibnamefont {Ohtomo}}, \bibinfo {author} {\bibfnamefont
  {K.}~\bibnamefont {Furusawa}}, \bibinfo {author} {\bibfnamefont
  {K.}~\bibnamefont {Hazu}}, \bibinfo {author} {\bibfnamefont {K.}~\bibnamefont
  {Hiramatsu}},\ and\ \bibinfo {author} {\bibfnamefont {A.}~\bibnamefont
  {Uedono}},\ }\bibfield  {title} {\enquote {\bibinfo {title} {Impacts of
  {{Si-doping}} and resultant cation vacancy formation on the luminescence
  dynamics for the near-band-edge emission of {{Al$_{0.6}$Ga$_{0.4}$N}} films
  grown on {{AlN}} templates by metalorganic vapor phase epitaxy},}\ }\href
  {https://doi.org/10.1063/1.4807906} {\bibfield  {journal} {\bibinfo
  {journal} {J. Appl. Phys.}\ }\textbf {\bibinfo {volume} {113}},\ \bibinfo
  {pages} {213506} (\bibinfo {year} {2013})}\BibitemShut {NoStop}%
\bibitem [{\citenamefont {Hyun~Kim}\ \emph {et~al.}(2023)\citenamefont
  {Hyun~Kim}, \citenamefont {Bagheri}, \citenamefont {Kirste}, \citenamefont
  {Reddy}, \citenamefont {Collazo},\ and\ \citenamefont
  {Sitar}}]{hyunkimTrackingPointDefects2023}%
  \BibitemOpen
  \bibfield  {author} {\bibinfo {author} {\bibfnamefont {J.}~\bibnamefont
  {Hyun~Kim}}, \bibinfo {author} {\bibfnamefont {P.}~\bibnamefont {Bagheri}},
  \bibinfo {author} {\bibfnamefont {R.}~\bibnamefont {Kirste}}, \bibinfo
  {author} {\bibfnamefont {P.}~\bibnamefont {Reddy}}, \bibinfo {author}
  {\bibfnamefont {R.}~\bibnamefont {Collazo}},\ and\ \bibinfo {author}
  {\bibfnamefont {Z.}~\bibnamefont {Sitar}},\ }\bibfield  {title} {\enquote
  {\bibinfo {title} {Tracking of {{Point Defects}} in the {{Full Compositional
  Range}} of {{AlGaN}} via {{Photoluminescence Spectroscopy}}},}\ }\href
  {https://doi.org/10.1002/pssa.202200390} {\bibfield  {journal} {\bibinfo
  {journal} {Phys. Status Solidi A}\ }\textbf {\bibinfo {volume} {220}},\
  \bibinfo {pages} {2200390} (\bibinfo {year} {2023})}\BibitemShut {NoStop}%
\bibitem [{\citenamefont {Prozheev}\ \emph {et~al.}(2020)\citenamefont
  {Prozheev}, \citenamefont {Mehnke}, \citenamefont {Wernicke}, \citenamefont
  {Kneissl},\ and\ \citenamefont
  {Tuomisto}}]{prozheevElectricalCompensationCation2020}%
  \BibitemOpen
  \bibfield  {author} {\bibinfo {author} {\bibfnamefont {I.}~\bibnamefont
  {Prozheev}}, \bibinfo {author} {\bibfnamefont {F.}~\bibnamefont {Mehnke}},
  \bibinfo {author} {\bibfnamefont {T.}~\bibnamefont {Wernicke}}, \bibinfo
  {author} {\bibfnamefont {M.}~\bibnamefont {Kneissl}},\ and\ \bibinfo {author}
  {\bibfnamefont {F.}~\bibnamefont {Tuomisto}},\ }\bibfield  {title} {\enquote
  {\bibinfo {title} {Electrical compensation and cation vacancies in {{Al}}
  rich {{Si-doped AlGaN}}},}\ }\href {https://doi.org/10.1063/5.0016494}
  {\bibfield  {journal} {\bibinfo  {journal} {Appl. Phys. Lett.}\ }\textbf
  {\bibinfo {volume} {117}},\ \bibinfo {pages} {142103} (\bibinfo {year}
  {2020})}\BibitemShut {NoStop}%
\bibitem [{\citenamefont {{van Deurzen}}\ \emph {et~al.}(2023)\citenamefont
  {{van Deurzen}}, \citenamefont {Singhal}, \citenamefont {Encomendero},
  \citenamefont {Pieczulewski}, \citenamefont {Chang}, \citenamefont {Cho},
  \citenamefont {Muller}, \citenamefont {Xing}, \citenamefont {Jena},
  \citenamefont {Brandt},\ and\ \citenamefont
  {L{\"a}hnemann}}]{vandeurzenExcitonicDeeplevelEmission2023}%
  \BibitemOpen
  \bibfield  {author} {\bibinfo {author} {\bibfnamefont {L.}~\bibnamefont {{van
  Deurzen}}}, \bibinfo {author} {\bibfnamefont {J.}~\bibnamefont {Singhal}},
  \bibinfo {author} {\bibfnamefont {J.}~\bibnamefont {Encomendero}}, \bibinfo
  {author} {\bibfnamefont {N.}~\bibnamefont {Pieczulewski}}, \bibinfo {author}
  {\bibfnamefont {C.~S.}\ \bibnamefont {Chang}}, \bibinfo {author}
  {\bibfnamefont {Y.}~\bibnamefont {Cho}}, \bibinfo {author} {\bibfnamefont
  {D.~A.}\ \bibnamefont {Muller}}, \bibinfo {author} {\bibfnamefont {H.~G.}\
  \bibnamefont {Xing}}, \bibinfo {author} {\bibfnamefont {D.}~\bibnamefont
  {Jena}}, \bibinfo {author} {\bibfnamefont {O.}~\bibnamefont {Brandt}},\ and\
  \bibinfo {author} {\bibfnamefont {J.}~\bibnamefont {L{\"a}hnemann}},\
  }\bibfield  {title} {\enquote {\bibinfo {title} {Excitonic and deep-level
  emission from {{N-}} and {{Al-polar}} homoepitaxial {{AlN}} grown by
  molecular beam epitaxy},}\ }\href {https://doi.org/10.1063/5.0158390}
  {\bibfield  {journal} {\bibinfo  {journal} {APL Mater.}\ }\textbf {\bibinfo
  {volume} {11}},\ \bibinfo {pages} {081109} (\bibinfo {year}
  {2023})}\BibitemShut {NoStop}%
\bibitem [{\citenamefont {Kaneko}\ \emph {et~al.}(2013)\citenamefont {Kaneko},
  \citenamefont {Okumura}, \citenamefont {Ishii}, \citenamefont {Funato},
  \citenamefont {Kawakami}, \citenamefont {Kimoto},\ and\ \citenamefont
  {Suda}}]{kanekoOpticalPropertiesHighly2013}%
  \BibitemOpen
  \bibfield  {author} {\bibinfo {author} {\bibfnamefont {M.}~\bibnamefont
  {Kaneko}}, \bibinfo {author} {\bibfnamefont {H.}~\bibnamefont {Okumura}},
  \bibinfo {author} {\bibfnamefont {R.}~\bibnamefont {Ishii}}, \bibinfo
  {author} {\bibfnamefont {M.}~\bibnamefont {Funato}}, \bibinfo {author}
  {\bibfnamefont {Y.}~\bibnamefont {Kawakami}}, \bibinfo {author}
  {\bibfnamefont {T.}~\bibnamefont {Kimoto}},\ and\ \bibinfo {author}
  {\bibfnamefont {J.}~\bibnamefont {Suda}},\ }\bibfield  {title} {\enquote
  {\bibinfo {title} {Optical {{Properties}} of {{Highly Strained AlN Coherently
  Grown}} on {{6H-SiC}}(0001)},}\ }\href
  {https://doi.org/10.7567/APEX.6.062604} {\bibfield  {journal} {\bibinfo
  {journal} {Appl. Phys. Express}\ }\textbf {\bibinfo {volume} {6}},\ \bibinfo
  {pages} {062604} (\bibinfo {year} {2013})}\BibitemShut {NoStop}%
\bibitem [{\citenamefont {{van Deurzen}}\ \emph {et~al.}(2022)\citenamefont
  {{van Deurzen}}, \citenamefont {Page}, \citenamefont {Protasenko},
  \citenamefont {Nomoto}, \citenamefont {Xing},\ and\ \citenamefont
  {Jena}}]{vandeurzenOpticallyPumpedDeepUV2022}%
  \BibitemOpen
  \bibfield  {author} {\bibinfo {author} {\bibfnamefont {L.}~\bibnamefont {{van
  Deurzen}}}, \bibinfo {author} {\bibfnamefont {R.}~\bibnamefont {Page}},
  \bibinfo {author} {\bibfnamefont {V.}~\bibnamefont {Protasenko}}, \bibinfo
  {author} {\bibfnamefont {K.}~\bibnamefont {Nomoto}}, \bibinfo {author}
  {\bibfnamefont {H.~G.}\ \bibnamefont {Xing}},\ and\ \bibinfo {author}
  {\bibfnamefont {D.}~\bibnamefont {Jena}},\ }\bibfield  {title} {\enquote
  {\bibinfo {title} {Optically pumped deep-{{UV}} multimode lasing in {{AlGaN}}
  double heterostructure grown by molecular beam homoepitaxy},}\ }\href
  {https://doi.org/10.1063/5.0085365} {\bibfield  {journal} {\bibinfo
  {journal} {AIP Adv.}\ }\textbf {\bibinfo {volume} {12}},\ \bibinfo {pages}
  {035023} (\bibinfo {year} {2022})}\BibitemShut {NoStop}%
\bibitem [{\citenamefont {Jmerik}, \citenamefont {Lutsenko},\ and\
  \citenamefont {Ivanov}(2013)}]{jmerikPlasmaassistedMolecularBeam2013}%
  \BibitemOpen
  \bibfield  {author} {\bibinfo {author} {\bibfnamefont {V.~N.}\ \bibnamefont
  {Jmerik}}, \bibinfo {author} {\bibfnamefont {E.~V.}\ \bibnamefont
  {Lutsenko}},\ and\ \bibinfo {author} {\bibfnamefont {S.~V.}\ \bibnamefont
  {Ivanov}},\ }\bibfield  {title} {\enquote {\bibinfo {title} {Plasma-assisted
  molecular beam epitaxy of {{AlGaN}} heterostructures for deep-ultraviolet
  optically pumped lasers},}\ }\href {https://doi.org/10.1002/pssa.201300006}
  {\bibfield  {journal} {\bibinfo  {journal} {physica status solidi (a)}\
  }\textbf {\bibinfo {volume} {210}},\ \bibinfo {pages} {439--450} (\bibinfo
  {year} {2013})}\BibitemShut {NoStop}%
\end{thebibliography}%

\end{document}